\documentclass[8.5pt,twoside,twocolumn]{article}
\usepackage[super,sort&compress,comma]{natbib} 
\usepackage{mhchem}
\usepackage{times,mathptm}
\usepackage{sectsty}
\usepackage{balance} 
\usepackage{amssymb}

\usepackage{graphicx} 
\usepackage{lastpage}
\usepackage[format=plain,justification=raggedright,singlelinecheck=false,%
  font=small,labelfont=bf,labelsep=space]{caption} 
\usepackage{fancyhdr}
\begin{document}

\thispagestyle{plain}
\fancypagestyle{plain}{
\fancyhead[L]{\includegraphics[height=8pt]{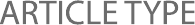}}
\fancyhead[C]{\hspace{-1cm}\includegraphics[height=20pt]{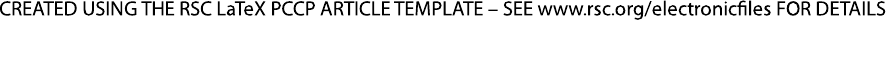}}
\fancyhead[R]{\includegraphics[height=10pt]{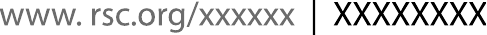}\vspace{-0.2cm}}
\renewcommand{\headrulewidth}{1pt}}
\renewcommand{\thefootnote}{\fnsymbol{footnote}}
\renewcommand\footnoterule{\vspace*{1pt}%
\hrule width 3.4in height 0.4pt \vspace*{5pt}} 
\setcounter{secnumdepth}{5}

\makeatletter 
\def\subsubsection{\@startsection{subsubsection}{3}{10pt}{-1.25ex plus -1ex minus -.1ex}{0ex plus 0ex}{\normalsize\bf}} 
\def\paragraph{\@startsection{paragraph}{4}{10pt}{-1.25ex plus -1ex minus -.1ex}{0ex plus 0ex}{\normalsize\textit}} 
\renewcommand\@biblabel[1]{#1}            
\renewcommand\@makefntext[1]%
{\noindent\makebox[0pt][r]{\@thefnmark\,}#1}
\makeatother 
\renewcommand{\figurename}{\small{Fig.}~}
\sectionfont{\large}
\subsectionfont{\normalsize} 

\fancyfoot{}
\fancyfoot[LO,RE]{\vspace{-7pt}\includegraphics[height=9pt]{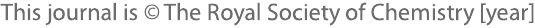}}
\fancyfoot[CO]{\vspace{-7.2pt}\hspace{12.2cm}\includegraphics{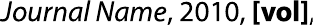}}
\fancyfoot[CE]{\vspace{-7.5pt}\hspace{-13.5cm}\includegraphics{headers/RF.pdf}}
\fancyfoot[RO]{\footnotesize{\sffamily{1--\pageref{LastPage} ~\textbar  \hspace{2pt}\thepage}}}
\fancyfoot[LE]{\footnotesize{\sffamily{\thepage~\textbar\hspace{3.45cm} 1--\pageref{LastPage}}}}
\fancyhead{}
\renewcommand{\headrulewidth}{1pt} 
\renewcommand{\footrulewidth}{1pt}
\setlength{\arrayrulewidth}{1pt}
\setlength{\columnsep}{6.5mm}
\setlength\bibsep{1pt}

\twocolumn[
  \begin{@twocolumnfalse}
\noindent\LARGE{\textbf{Schematic Mode Coupling Theory of Glass Rheology: \\ Single and Double Step Strains}}
\vspace{0.6cm}

\noindent\large{\textbf{Th. Voigtmann,\textit{$^{a,b,c}$} J M Brader,\textit{$^d$} M Fuchs,\textit{$^a$}, and M E Cates$^\ast$\textit{$^{e}$}}}\vspace{0.5cm}

\noindent\textit{\small{\textbf{Received Xth XXXXXXXXXX 20XX, Accepted Xth XXXXXXXXX 20XX\newline
First published on the web Xth XXXXXXXXXX 20XX}}}

\noindent \textbf{\small{DOI: 10.1039/b000000x}}
\vspace{0.6cm}

\noindent \normalsize{Mode coupling theory (MCT) has had notable successes in addressing the rheology of hard-sphere colloidal glasses, and also soft colloidal glasses such as star-polymers. 
Here, we explore the properties of a recently developed MCT-based schematic constitutive equation under idealized experimental protocols involving single and then double step strains. We find strong deviations from expectations based on factorable, BKZ-type constitutive models. Specifically, a nonvanishing stress remains long after the application of two equal and opposite step strains; this residual stress is a signature of plastic deformation. We also discuss the distinction between hypothetically instantaneous step strains and fast ramps. These are not generally equivalent in our MCT approach, with the latter more likely to capture the physics of experimental `step' strains. The distinction points to the different role played by reversible anelastic, and irreversible plastic rearrangements.}
\vspace{0.5cm}
 \end{@twocolumnfalse}
  ]

\section{Introduction}


\footnotetext{\textit{$^{a}$~Fachbereich Physik, Universit\"at Konstanz, 78457 Konstanz, Germany. }}
\footnotetext{\textit{$^{b}$~Zukunftskolleg, Universit\"at Konstanz, 78457 Konstanz, Germany. }}
\footnotetext{\textit{$^{c}$~Institut f\"ur Materialphysik im Weltraum, Deutsches Zentrum f\"ur Luft- und Raumfahrt (DLR), 51170 K\"oln, Germany. }}
\footnotetext{\textit{$^{d}$~Department of Physics, University of Fribourg, CH-1700 Fribourg, Switzerland. }}
\footnotetext{\textit{$^{e}$~SUPA, School of Physics and Astronomy, University of Edinburgh, Mayfield Road, Edinburgh, EH9 3JZ, Scotland. }}


 

The nonlinear flow and deformation of a material can reveal a wealth of information on its internal relaxation mechanisms. The theoretical prediction
and understanding of this nonlinear rheological response is, however, a formidable task.
Ideally, one aims to construct a constitutive equation, i.e., a relation
that predicts the stress tensor at any time $t$ as a functional of the deformation history \cite{LarsonCE}. 
The task is particularly relevant for systems with large internal relaxation
times, where the regime of nonlinear response is easily reached by experiment. Such systems include entangled polymers, in melts and solutions, which have long been a prime area of study in nonlinear rheology \cite{LarsonCE}. Recently however, the case of arrested colloidal states, comprising either hard or soft spherical particles in suspension at high density, has also received considerable experimental attention \cite{Petekidis2002,Haw2004,Crassous2006,Pham2006,Besseling2007,Isa2007,Crassous2008,Pham2008,Zausch2008,Siebenbuerger2009,Negi2009,Besseling2010,Bandyopadhyay2010,Negi2010,Laurati2011,Willenbacher2011}. These dense colloidal systems have relaxation times that are very long (like polymers) or even, within the colloidal glass phase, effectively infinite. The non-ergodicity of the glass poses special problems to rheological theory, and is the cause of important new features such as a yield stress. The flow of arrested colloids has recently been addressed via phenomenological theory \cite{Derec2000,Coussot2002}, mesoscopic models \cite{SGR,Sollich1998,STZ1,STZreview} and  first principles methods such as mode-coupling theory (MCT) \cite{FuchsCates,Brader2007,Brader2008}. An interesting recent development has been the formulation of an MCT-inspired schematic constitutive model capable of addressing arbitrary time dependent flows \cite{Brader2009}. This is significantly simpler than the full microscopic MCT, but appears to capture its main features. In this paper we explore some predictions of this schematic MCT approach, focusing on single and double step strains.

Double step strain deformations are a commonly used protocol for rheology and
materials testing. For simplicity, we will only consider shear deformations here, denoting shear strain by $\gamma(t)$ and
the relevant off-diagonal element of the stress tensor by $\sigma(t)$.
In such an experiment one first applies a (usually large) shear strain $\gamma_0$
at, say, time $t=0$. Although in theory often idealized as a discontinuous step, this will usually take some small, finite time $\delta t$ during
which a strain ramp is applied. After a waiting time $\Delta t$, a second
strain $\gamma_1$ is applied, and the subsequent stress relaxation is
measured. Of course, the response to a single step already provides information on
the relaxation processes of the material. However, by tuning $\Delta t$ to be
comparable with the time scale of those processes, the double-step strain
response can offer a more sensitive probe of the dynamics.

This holds in particular for the case of reversing double step strains, where the
steps act in opposite directions: $\gamma_0>0$ but $\gamma_1<0$.
The case where the total strain imposed on the system vanishes,
$\gamma_T=\gamma_0+\gamma_1 = 0$, allows an especially clear assessment of the deformation
energy and its recovery. Consider a nonlinear but elastic solid undergoing
a single step strain: at long times, a residual stress will remain
that relates to the strain energy stored in the system.
Adding a second step strain, the residual stresses created by both
will combine in a nonlinear fashion. However, for a purely elastic material, or indeed any viscoelastic material with a unique equilibrium configuration (so-called `anelastic' materials), the residual stress must vanish eventually whenever $\gamma_T=0$. The residual stress will also vanish eventually in any viscoelastic liquid (whose longest relaxation time is by definition finite, so all stresses relax). 

In view of those comments it is perhaps unsurprising that the ultimate vanishing of the stress $\sigma(t=\infty,\gamma_T=0)$ after exactly reversed double-step strain is predicted by most established constitutive models of the nonlinear rheological response in viscoelastic solids and fluids. 
In particular, the result $\sigma(\infty,0) = 0$ is unavoidable for any constitutive model in which the stress may be written as:
\begin{equation}\label{larson}
  \sigma(t)=\int_{-\infty}^t \psi(t-t';\gamma_{tt'})\gamma_{tt'}\,dt'
\end{equation}
where $\gamma_{tt'}$ is the accumulated shear strain between times $t$ and
$t'$. This form is (for simple shear flows) equivalent to the so-called BKZ
(or Kaye-BKZ or K-BKZ) equation \cite{BKZ,Kaye}. It approximates the response
to a general flow by a nonlinear superposition (called Boltzmann superposition
principle\cite{LarsonCE}) whose kernel is fixed by the stress response $\psi(t,\gamma)$ to a
single step strain of amplitude $\gamma$. For double step strain,
Eq.~\eqref{larson} yields
\begin{equation}
  \sigma_\text{BKZ}^{(2)}(t)
  =\sigma^{(1)}_{\gamma_0+\gamma_1}(t)
  -\sigma^{(1)}_{\gamma_1}(t)
  +\sigma^{(1)}_{\gamma_1}(t-\Delta t)
  \label{bkz}
\end{equation}
for times after the second step.
Here $\sigma^{(1)}(t)$ denotes the response after a single step,
and $\sigma^{(2)}(t)$ the response after two steps separated by $\Delta t$,
for $t>\Delta t$.
As $t\to\infty$, BKZ-type models predict that the response to the
double step approaches that of a single step of combined magnitude.

On the other hand, if in a solid material an irreversible plastic deformations arises between the initially imposed strain and its later exact reversal, the residual stress $\sigma(\infty,0)$ has no reason to vanish. Consider for example
a Bingham plastic, a hypothetical material of modulus $G$ showing linear
elasticity up to strain $\gamma_y$ and pure plasticity beyond.
A step strain of (say) $3\gamma_y/2$ gives a recoverable strain $\gamma_y$;
reversing the step takes this to $-\gamma_y/2$ so that
$\sigma(\infty) = -G\gamma_y/2$. With this in mind, one should not be
surprised if well founded constitutive equations for colloidal glasses, or
indeed other materials where plasticity plays a major role, end up looking
considerably more complicated than Eq. \ref{larson}. 

Most constitutive equations to date for viscoelastic liquids and solids have been conceived within the framework of continuum elasticity theory or fluid
mechanics. A notable exception that starts from first principles equations of motion is the Doi-Edwards theory of polymer melts and solutions, and its
later refinements \cite{DoiEdwards}.
Recently, comparable steps towards a constitutive model for dense colloidal suspensions
have been undertaken through an extension of the mode-coupling theory
of the glass transition based on an integration-through transients
formalism (ITT-MCT) \cite{FuchsCates,Fuchs2009,Brader2007}. Starting from the Smoluchowski equation including
applied flow, and involving a set of approximations tailored to capture
the collective structural-relaxation dynamics of dense liquids, equations
are derived for the transient nonequilibrium density correlation functions.
These in turn determine the stress tensor through a suitably approximated
generalization of a Green-Kubo relation. The memory effects incorporated in
MCT imply a nonlinear dependence on the whole strain history.

ITT-MCT is a microscopic theory. It takes the static structure factor of the quiescent system as input, and can therefore predict, in principle, rheological differences between colloidal systems with different interactions (hard and soft colloids;  colloids with or without attractive interactions, etc.). Unfortunately however, to actually compute its predictions for all but the simplest flows remains a technically challenging task \cite{Brader2008,Weysser2009,FuchsEPJE}.
To understand better the
qualitative and generic features of the resulting constitutive equations, a schematic model
has been proposed that aims to incorporate the essential mathematics
of the microscopic equations while reducing their complexity \cite{Brader2009}.

In this paper, we analyse this schematic MCT model for the case of single and double step strains, focussing in particular on the residual stresses following double
step strain deformations. We include a careful discussion of how exactly step strains should be implemented within the MCT approach, bearing in mind that in the laboratory `step strains' are not genuinely discontinuous, but merely rapid on the time scale of structural relaxtion.

\section{Schematic MCT}

We now recapitulate the equations defining the schematic model for
colloidal rheology; for more details and a discussion of the relations with
microscopic ITT-MCT, see Ref.~\citep{Brader2009}. For pure shear with
a given shear rate $\dot\gamma(t)$,
\begin{equation}\label{gk}
  \sigma(t)=\int_{-\infty}^t\dot\gamma(t')G(t,t';[\gamma(t')])\,dt'\,.
\end{equation}
$G(t,t';[\gamma(t')])$ is a generalized dynamical shear modulus. In the nonlinear
response regime it depends functionally on the strain history $\gamma(t'<t)$ through the accumulated strain
evaluated at various times (see below). Outside the steady-state regime,
it depends (as does any two-time correlation function) on times $t$ and $t'$ separately, not just the interval $t-t'$.
In MCT, the generalized modulus is expressed through transient density correlation functions.
For the schematic model, we write
\begin{equation} \label{GDEF}
  G(t,t';[\gamma(t')])=v_\sigma\phi(t,t')^2\,.
\end{equation}
Here the parameter $v_\sigma$ sets the scale of the shear modulus; for instance
the static linear modulus in the quiescent glass is $G_\infty=v_\sigma f_0^2$
where $f_0$ is the nonergodicity parameter of the idealized glass, given by the
long-time limit of the density correlation function (see below).
Here we set $v_\sigma=100$ which gives roughly the correct modulus, in units of $k_BT/d^3$ with $d$ the colloidal diameter, for hard spheres close to the glass transition \cite{Brader2009}.
The density correlator $\phi(t,t')$, whose dependence on strain history is no longer written explicitly but still present, obeys a generalization
of the Mori-Zwanzig memory equation
\begin{equation}\label{mz}
  \tau_0\partial_t\phi(t,t')+\phi(t,t')
  +\int_{t'}^th_{tt'}h_{tt''}m(t,t'')\partial_{t''}\phi(t'',t')\,dt''=0\,.
\end{equation}
Here $\tau_0$ is a microscopic (Brownian) relaxation time; for our numerical
work we choose time units such that $\tau_0=1$.
Also, $m(t,t'')$ is a memory kernel describing the slow
dynamics close to the glass transition. Following MCT precepts, this is approximated schematically
as a nonlinear functional of the correlator itself. Specifically, following the form adopted in the so-called
F12 model, we have
\begin{equation}\label{f12}
  m(t,t')\equiv m[\phi(t,t')]=v_1\phi(t,t')+v_2\phi(t,t')^2\,.
\end{equation}
Parameters $v_1$ and $v_2$ determine the state point of the model, controlling in combination the distance from the glass transition and its strength (which will depend in principle on the nature of the colloidal interactions). In the quiescent system there accordingly appears a line of critical coupling coefficients $(v_1^c,v_2^c)$ that
separates liquid-like responses for small coupling strengths from solid-like
responses at large coupling strengths. The solid phase is an idealized glass, in which (within MCT) the longest relaxation time is not merely large, but infinite. The glass is therefore
characterized by a nonergodic contribution to the correlation
functions: in a quiescent system possessing an equilibrium Boltzmann distribution, $\phi(t,t')=\phi_0(t-t')\to f_0>0$ for
$t-t'\to\infty$.
The nonergodicity parameter $f_0$ of the model is the largest non-negative
solution of $1/(1-f)=1+m[f]$.
At the glass transition, the long-time quiescent limit jumps discontinuously from $f_0=0$ to
$f_0 = f^c$. For our numerical calculations, we follow convention \cite{Brader2009} and set $v_2^c=2$ which implies
$v_1^c=2(\sqrt2-1)$ and $f_c=1-1/\sqrt2$. There remains
a single state parameter $\varepsilon$ measuring how far the sample lies from its glass transition:
$v_1=v_1^c+\varepsilon/(\sqrt2-1)$, $v_2=v_2^c$.
Ergodic liquid states have $\varepsilon<0$, ideal glass states have
$\varepsilon>0$.

The ``damping functions'' $h$ that appear in Eq.~\eqref{mz} incorporate the nonlinear reduction of
memory effects caused by strain. Strain can break the cage around each particle by forcing rearrangement of its neighbours; this cage is the cause of the memory effects which lead to permanent trapping of the particle in the quiescent glass state.
The damping functions are decreasing functions of increasing
strain, and modeled schematically as\cite{fingernote}
\begin{equation}\label{hfunc}
  h_{tt'}=1/[1+(\gamma_{tt'}/\gamma_c)^2]
\end{equation}
where $\gamma_{tt'}=\int_{t'}^t\dot\gamma(\tau)\,d\tau$ is the accumulated shear strain between times $t'$ and $t$.
Here, $\gamma_c$ is a parameter controlling the critical strain that is
sufficient to break cages. We set $\gamma_c=1/10$ to match with
experimental findings that this deformation is on the scale of a typical
cage size, given by the Lindemann criterion \cite{Brader2009,Zausch2008}.

\section{Step Strains within MCT}
A step strain of amplitude $\gamma$ can, within MCT, be represented mathematically in two distinct ways. The obvious method is simply to set $\dot\gamma(t) = \gamma\delta(t)$ in the equations presented above. It was the route chosen in Refs.~\cite{Brader2007,Brader2009}. This results in an instantaneous, anelastic (or nonlinearly elastic) deformation which has no immediate effect on the memory functions: these only start to alter once relaxation of the newly applied step deformation begins.

The second route is to consider the same total strain to be applied at a finite ramp rate. Intriguingly, this is not equivalent to the delta function even if the ramp rate is later taken to be (within the schematic model) infinitely fast. The cause of this distinction is subtle, but clear. It stems from the fact that in MCT the structural relaxation time in sustained shear is the inverse of the shear rate itself. Thus if a large strain $\gamma$ is applied at any finite rate, this leads to a plastic deformation whose cumulative value does not vanish as that rate increases. By the end of the ramp, if $\gamma$ is large enough, all memory of the previous flow history (including the early part of the ramp itself) has been erased, which is not the case in the delta-function approach. We will see that consequences of this for single step strains, as considered in previous work \cite{Brader2007,Brader2009} are relatively minor unless $\gamma\gg \gamma_c$.  However the additional memory loss becomes important in double step strain, where the response after a second step depends strongly on how much this erases memory of the first. Here we will find qualitative differences between the two approaches.

In deciding which representation is more appropriate, one must bear in mind that ITT-MCT, on which the schematic model is based, explicitly restricts shear rates to the range of small (bare) Peclet number, $\dot\gamma\tau_0\ll1$. Thus taking the limit of a fast finite ramp is, in principle, just as questionable as adopting the delta function. However, as will be seen from the results that follow, we find no sign of any singular behaviour on taking that finite ramp rate to infinity: indeed the results are broadly similar to those found for $\dot\gamma\tau_0\simeq 1$.
Thus the fast ramp limit may offer a convenient model of ramps that are experimentally step-like (fast compared to the observed quiescent structural relaxation times of the material) yet still slow on the microscopic timescale set by $\tau_0$. The delta function approach, for the reasons given above, does not represent this case. It might possibly capture instead ramps where $\tau_0\dot\gamma\gg 1$ (if in consequence plastic rearrangement is not possible); however, as stated already, MCT is not to be trusted in that regime. Moreover, it is questionable whether conventional rheological experiments on colloidal materials can involve step strains that are fast enough to approach this limit of high bare Peclet number during the imposition of the step.

For these various reasons we currently prefer the finite ramp approach as a predictive tool for experimental colloid rheology, and develop this next. After that, we summarize for comparison the results of the delta-function approach, before giving our conclusions.

\section{`Step' Strains Comprising Finite Ramps}

Here we consider single and double step-strain experiments in which the `steps' represent strain
ramps of finite width $\delta t$. For simplicity, we assume a constant
shear rate for each of these ramps, so that
\begin{equation}
  \dot\gamma(t)=\begin{cases}
  \dot\gamma_0 & \text{for $0<t<\delta t$}\\
  \dot\gamma_1 & \text{for $\Delta t+\delta t<t<\Delta t+2\delta t$}\\
  0 & \text{else}
  \end{cases}
\end{equation}
The shear rates are given by $\dot\gamma_i=\gamma_i/\delta t$ for $i=0,1$.

In writing the MCT equations for this shear-rate protocol, one
is led to consider nine different correlators (seven if the
two shear rates are of equal magnitude) for the calculation of
$\sigma(t)$. This can be seen from splitting
the domain $(t'<t)$ of the $(t,t')$ plane into regions bounded by
the discontinuities along the $t$- and $t'$-axes, and realizing that
$(t'<0)$ is not needed in evaluating Eq.~\eqref{gk}.

The resulting equations are defered to the Appendix. Let us first highlight
some important features for the case of a single step strain ($\dot\gamma_1=0$). In that case, only
three correlators are needed. Two of these are well known from previous work \cite{Brader2009,FuchsFaraday}: the quiescent equilibrium one $\phi_0(t-t')$
(needed for $t\ge t'>\delta t$), and the steady-state $\phi_\text{ss}^{(\dot\gamma_0)}(t-t')$ (for
$\delta t>t\ge t'$). The third is a new two-step correlator which we denote $\phi^{(10)}(t,t')$; this is needed
for $t>\delta t>t'$. The corresponding $G^{(10)}(t,t')$ enters the
Green-Kubo integral,
\begin{equation}\label{barsigma1-ramp}
  \bar\sigma^{(1)}(t)=\int_0^{\delta t}\dot\gamma_0G^{(10)}(t,t')\,dt'
  =\gamma_0\int_0^1\bar G^{(10)}(t,x)\,dx
\end{equation}
where we use the overbar on $\bar\sigma$ as a reminder that the ramp rate is finite and the superscript $(1)$ to label the single-step response. We have also defined $\bar G^{(10)}(t,x)=G^{(10)}(t,t')$ when $t'=x\,\delta t$.
The equation of motion for the correlator reads
\begin{multline}\label{phi10}
  \tau_0\partial_t\bar\phi^{(10)}(t,x)+\bar\phi^{(10)}(t,x)\\
  +h[\gamma_0(1-x)]\int_x^1h[\gamma_0(1-y)]m[\bar\phi^{(10)}(t,y)]
  \partial_y\phi_\text{ss}((y-x)\delta t;\dot\gamma_0)\,dy\\
  +h[\gamma_0(1-x)]\int_{\delta t}^tm[\phi_0(t-t'')]\partial_{t''}
  \bar\phi^{(10)}(t'',x)\,dt''=0
\end{multline}
An analysis of the behavior of the solutions to this equation at arbitrary
$\delta t$ is intricate. Setting aside for a moment the restriction to
small Peclet numbers, the limit $\delta t\to0$ is worth discussing.
Note that the first integral then vanishes as $\mathcal O(\delta t)$.
Equation~\eqref{phi10} thus turns into a one-parameter set of equations
for $x\in[0,1]$, as stated in the Appendix (Eq.~\eqref{eomphi10}).
As explained there,
each of these is independently solved by a single-step strain
correlator $\phi^{(1)}_{2\gamma_0(1-x)}(t)$, where
$\phi^{(1)}_{\gamma_0}(t)$
obeys
\begin{equation}\label{phi1}
  \tau_0\partial_t\phi_{\gamma_0}^{(1)}(t)+\phi_{\gamma_0}^{(1)}(t)
  \\
  +h(\gamma_0/2)\int_0^tm[\phi_0(t-t')]
  \partial_{t'}\phi_{\gamma_0}^{(1)}(t')\,dt'=0\,.
\end{equation}
We will meet
$\phi^{(1)}_{\gamma_0}(t)$ again below when we address delta-function steps; in fact, it is the correlator found there by setting $\dot\gamma(t) = \gamma_0\delta(t)$ in our schematic equations of motion.
We thus get for $t>\delta t$
\begin{multline}\label{dt0-ramp-single}
  \lim_{\delta t\to0}
  \bar\sigma^{(1)}_{\gamma_0}(t)
  =\gamma_0\int_0^1dx\,G^{(1)}_{2\gamma_0(1-x)}(t-x\delta t)
  \\
  =\int_0^{\gamma_0}d\eta\,G^{(1)}_{2\eta}(t)
\end{multline}
where 
$G^{(1)}_{\gamma_0}(t)$ is related to the correlator $\phi^{(1)}_{\gamma_0}(t)$
via Eq.~\ref{GDEF}.

The generalization of Eq.~\eqref{barsigma1-ramp} to two strain ramps reads
\begin{multline}\label{barsigma2-ramp}
  \bar\sigma^{(2)}(t)=\int_0^{\delta t}\dot\gamma_0
  G^{(30)}_{\gamma_0,\gamma_1}(t,t')\,dt'
  \\
  +\int_{0}^{\delta t}\dot\gamma_1
  G^{(10)}_{\gamma_1}(t-\Delta t-\delta t,t')\,dt'
\end{multline}
where we have taken the ramps of equal length $\delta t$, and the expression
holds for $t>\Delta t+2\delta t$.
Here $G^{(10)}_{\gamma_1}$ is again the shear modulus appearing in the
single-strain ramp discussed above, but evaluated for the strain $\gamma_1$.
The equation for $G^{(30)}$ is given in the Appendix.
If one of $\delta t$ or $\Delta t$ remains finite, Eq.~\eqref{barsigma2-ramp} does not
reduce to a form that can be expressed through single-step response
functions. Only if both $\delta t\to0$ and $\Delta t\to0$, a simple
expression can be obtained in analogy to Eq.~\eqref{dt0-ramp-single},
\begin{multline}\label{dt0-ramp}
  \lim_{\delta t\to0,\Delta t\to0}\bar\sigma^{(2)}(t)
  =\gamma_0\int_0^1dx\,G^{(1)}_{2\gamma_0(1-x)+2\gamma_1}(t-x\delta t)\\
  +\gamma_1\int_0^1dx\,G^{(1)}_{2\gamma_1(1-x)}(t-\Delta t-\delta t-x\delta t)
  \\
  =\int_{\gamma_1}^{\gamma_0+\gamma_1}d\eta\,G^{(1)}_{2\eta}(t)
  +\int_0^{\gamma_1}d\eta\,G^{(1)}_{2\eta}(t-\Delta t)
  \\
  =\bar\sigma^{(1)}_{\gamma_0+\gamma_1}(t)-\bar\sigma^{(1)}_{\gamma_1}(t)
  +\bar\sigma^{(1)}_{\gamma_1}(t-\Delta t)
\end{multline}
(This is seen by substituting $\eta=\gamma_0(1-x)+\gamma_1$
in the first, and $\eta=\gamma_1(1-x)$ in the second, integral.)
In the last equality, we recognize the BKZ prediction, Eq.~\eqref{bkz}.
It is obtained from the ramp-MCT model only in the double limit
$\delta t\to0$ and $\Delta t\to0$ (taken in either order), because in
this case, cross-correlations that appear for finite $\Delta t$ become
irrelevant.

\subsection{Results for Ramp Approach}

We now discuss numerical results found from Eq.~\eqref{mz} and \eqref{f12} using the ramp approach to  step strains as detailed above. Adapting a standard algorithm for solving the steady-state
MCT equations, the equations of motion were solved by separately considering
the correlation functions defined in the various $(t,t')$ regions between
the discontinuities in $\dot\gamma$ (sketched in the Appendix). A large time window is covered in
the numerics by repeatedly doubling the initial discretization step size.
This implicitly assumes that all correlation functions become slowly varying
if the difference of their time arguments is large. This is well established
for the quiescent and steady-state correlators of MCT. Since the equations
of motion share the same structure, we believe it to hold also for the
various correlation functions discussed above (and have not yet found any exceptions to this numerically).

\begin{figure}
\includegraphics[width=.9\linewidth]{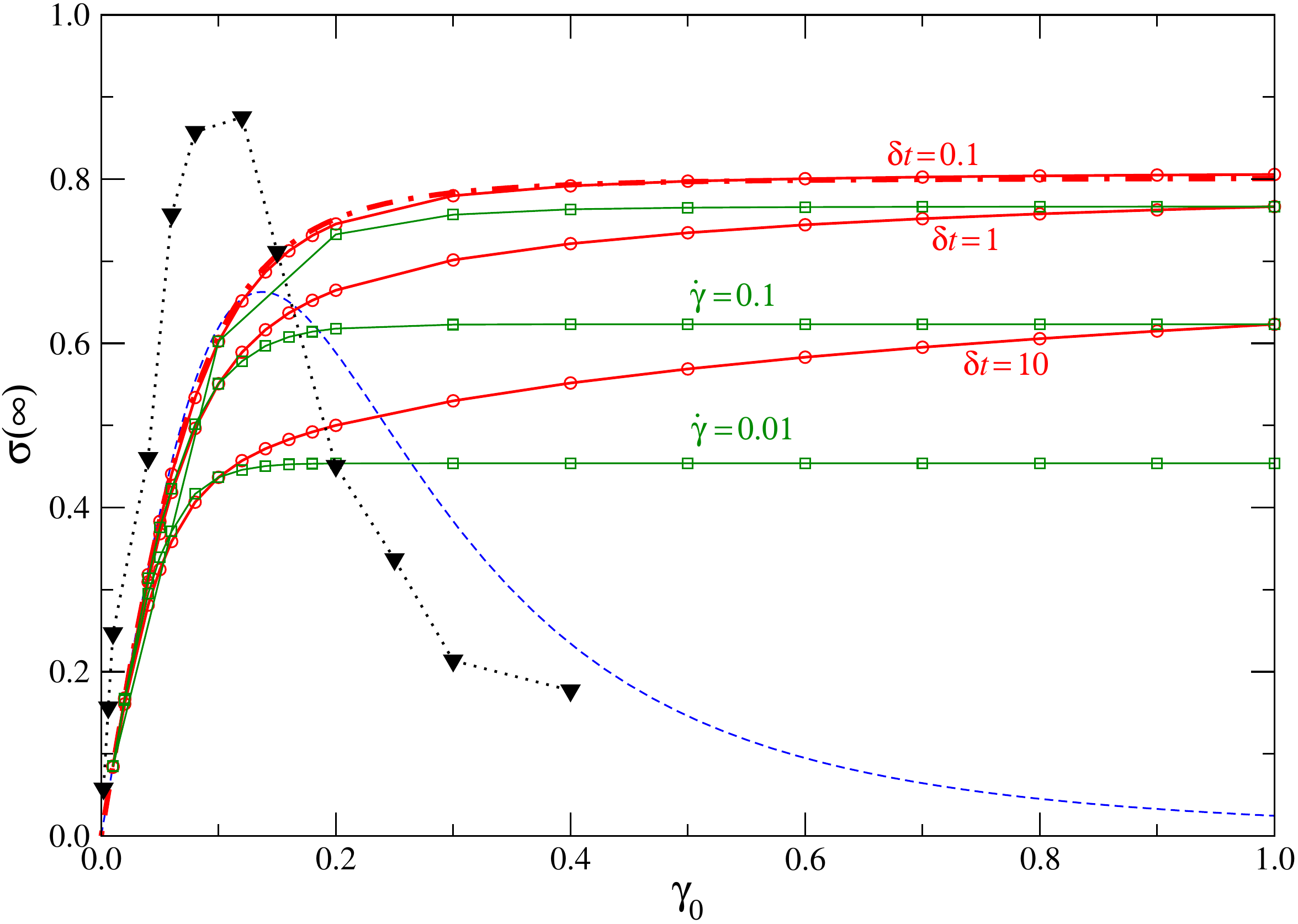}
\caption{\label{stepstrain-dtsiginf}
  $\bar\sigma(\infty)$ after step strain with amplitude $\gamma_0$,
  for the F12 model at the glass transition, $\varepsilon=0$,
   following a constant-shear-rate ramp
  of width $\delta t$ (dash-dotted: analytical result for $\delta t\to0$;
  circles: for $\delta t$ as labeled; squares: for fixed $\dot\gamma_0
  =\gamma_0/\delta t$ as labeled).
The result $\sigma(\infty)$ predicted using a delta-function step strain is also shown
  (dashed curve).}
\end{figure}

We start with the case of a single step strain.
In the ideal glass phase ($\varepsilon > 0$), the quiescent-state correlation function does not decay
but exhibits a finite long-time limit $f_0$. As a consequence, 
the memory kernel also does not decay, $m_0(t\to\infty)=m_0[f_0]=f_0/(1-f_0)>0$.
Moreover a step strain is unable to melt the glass on a permanent basis, so that a finite long-time limit exists also for the memory function and resulting correlator $\phi^{(10)}(t,t')$ needed to compute the stress response to such a step, via Eqs.~\eqref{GDEF} and \eqref{dt0-ramp-single}. Accordingly, that stress response does not decay to zero but to a finite value $\bar\sigma(\infty)$. 
In the limit of short ramps, $\delta t\to 0$, the result is found as 
\begin{equation}\label{siginf-ramp}
  \lim_{\delta t\to0}\bar\sigma^{(1)}(\infty)=v_\sigma\int_0^1\gamma_0dx\,
  \left(\frac{h[\gamma_0(1-x)]m_0[f_0]}{1+h[\gamma_0(1-x)]m_0[f_0]}\right)^2
\end{equation}
For the choice of the damping function in Eq.~\eqref{hfunc}, this integral can
be evaluated; for $\gamma_0\to\infty$ it gives
$\lim_{\delta t\to0}\bar\sigma^{(1)}(\infty)
\to v_\sigma(\pi/4)\gamma_cf_0^2>0$, i.e., the large-$\gamma_0$ behavior saturates to a constant. The dash-dotted curve in Fig.~\ref{stepstrain-dtsiginf}
shows the analytical result of Eq.~\eqref{siginf-ramp}
together with numerical results for finite
$\delta t$ (solid lines with circle symbols). In each case, a linear regime
extends up to $\gamma_0\approx\gamma_c$, indicating the response of an
elastic solid, $\sigma=G_\infty\gamma_0$, where $G_\infty$ is the glass
plateau modulus. After that, a sublinear regime indicates plastic deformation.

The large-$\gamma_0$ asymptote can be understood by recognizing that
in Eq.~\eqref{siginf-ramp}, $\gamma_0h[\gamma_0(1-x)]^2$ approaches an even-function
representation of $(\pi/2)\gamma_c\delta(1-x)$ as $\gamma_0\to\infty$.
The integral is therefore dominated by $x=1$. In other words, as
$\delta t\to0$, the shear rate $\dot\gamma_0\to\infty$, and the
shear-induced loss of memory is perfect for all $t'<\delta t$. Hence the
the $t>\delta t$ response is unaffected by the imposed shear except 
for a narrow window at
the time $t'=\delta t$ where the shear rate is just being switched off.
This is in contrast to the $\delta$-step model, explored (for a single step) in previous full \cite{Brader2007} and  schematic
\cite{Brader2009,Brader2009note} MCT calculations, whose result for the residual stress $\sigma(\infty)$ is also shown in Fig.\ref{stepstrain-dtsiginf}
(dashed line).
Here, the limit $\gamma_0\to\infty$ yields zero, approached as
$\propto\gamma_0h(\gamma_0/2)^2\sim1/\gamma_0^3$ in our case.

As previously discussed, the finite-ramp approach becomes questionable when $\dot\gamma_0\to\infty$,
since the microscopic derivation of ITT-MCT assumed small Peclet numbers,
$\dot\gamma_0\tau_0\ll1$. However, applying the same strain in
ramps with lower shear rates shows that the fast ramp limit is nonsingular; this is shown by the green
squares in Fig.~\ref{stepstrain-dtsiginf}. For these slower ramps, after an elastic
regime for $\gamma_0\lesssim\gamma_c$, a constant asymptote is again quickly
approached for large $\gamma_0$. This is because the
dynamics in the interval $[0,\delta t]$ is governed by the transient
steady-state correlator, which decays on a time scale
$\tau_{\dot\gamma_0}\sim \gamma_c/\dot\gamma_0$ and becomes a function only
of $\gamma_0 = \dot\gamma_0t$ only at long times. For $\gamma_0\gg\gamma_c$,
the ramp time is always long compared to this relaxation,
$\delta t\gg\tau_{\dot\gamma_0}$. The integral in Eq.~\eqref{barsigma1-ramp}
then extends over the full steady-state correlator and no longer scales
with $\gamma_0$, leading to a $\gamma_0$-independent constant. In consequence, for $\gamma_0\gg \gamma_c$, the residual stress
$\bar\sigma^{(1)}(\infty)$ is the same as would arise after cessation
of steady shear. (Note that, within our schematic MCT, the latter quantity attains a finite limit as the shear rate diverges.)

\begin{figure}
\includegraphics[width=.9\linewidth]{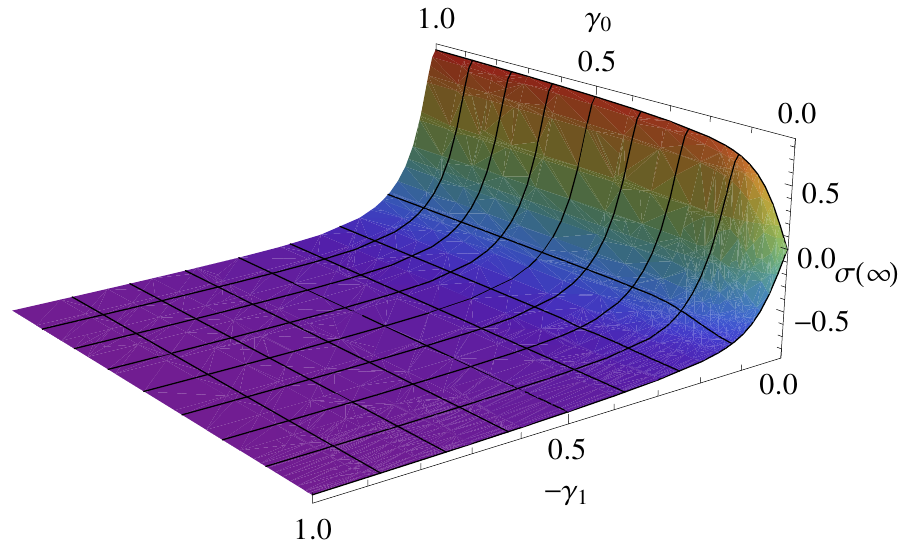}\\
\includegraphics[width=.9\linewidth]{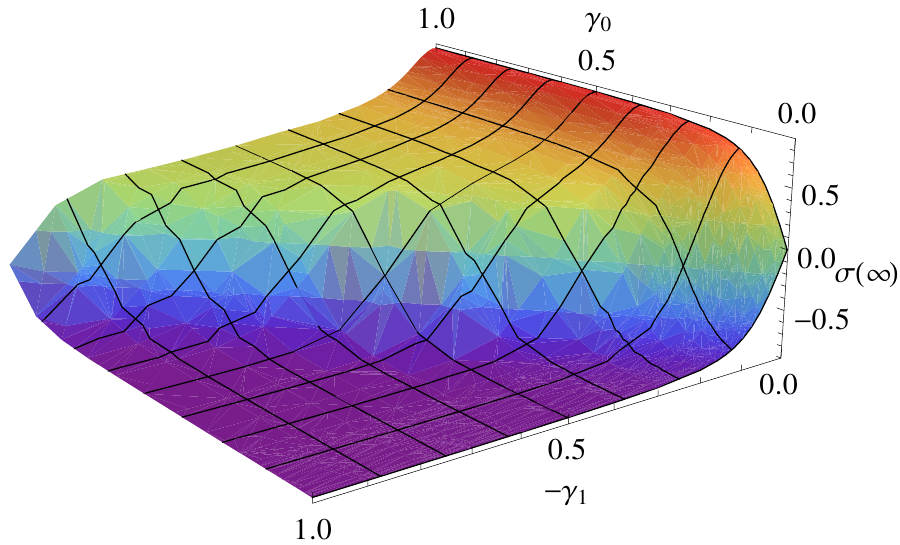}
\caption{\label{dstepx-ramp}
  Infinite-time stress plateau $\bar\sigma(\infty)$ after reversing double step
  strains in the glass (F12 model, $\varepsilon=10^{-4}$),
  as a function of the first and second
  step sizes, $\gamma_0$ positive and $\gamma_1$ negative.
  Top: Time between the strains $\Delta t=10$, with strain ramps of
  width $\delta t = 1$.
  Bottom: $\Delta t=\delta t=0.1$.
}
\end{figure}

We now turn to the case of a double step strain, where the residual stress is due
to a nonlinear combination of the two single steps. Fig.~\ref{dstepx-ramp}
shows results for the resulting residual stress $\bar\sigma^{(2)}(\infty)$ after a reversing step strains,
where $\gamma_0>0$ but $\gamma_1<0$, as a function of the two strain
amplitudes. 

Figure~\ref{dstepx-ramp} shows a remarkable difference of the MCT model to
the expectation from BKZ-type constitutive equations when $\Delta t$ is
large enough. In the latter,
the nonlinear superposition principle, Eq.~\eqref{larson}, implies that
the response to an arbitrary strain history can be constructed from those
to single step-strain experiments. In particular, Eq.~\eqref{bkz} implies 
$\sigma^{(2)}(\infty)=\sigma^{(1)}_{\gamma_0+\gamma_1}(\infty)$, so that
the plot in Fig.~\ref{dstepx-ramp} should be inflection-symmetric around
the line $\gamma_0=-\gamma_1$. As a consequence, applying two large strains
in opposite directions but such that the total strain $\gamma>0$,
results in a positive residual stress according to BKZ-type models,
no matter how large the waiting time between the strains.
In the ramp-step MCT model, a physically more plausible prediction emerges.
Whereas for $\Delta t\to0$ and $\delta t\to0$,
the nonlinear superposition principle is formally recovered, cf.\
Eq.~\eqref{dt0-ramp}, for $\Delta t\gg\tau_0$, the residual stress
$\bar\sigma(\infty)$ is not a function
of $\gamma_0+\gamma_1$ alone. 
Instead it is seen from Fig.~\ref{dstepx-ramp} that the result becomes
independent of $\gamma_0$ if $|\gamma_1|\gtrsim \gamma_c$. 
The physics was already described above for a single step: when the strain amplitude of the second step is large, $|\dot\gamma_1|\delta t\gg\gamma_c$
holds. The system then approaches a steady-state flow during the second strain ramp,
which erases all memory of the past deformation history, including everything related to the first ramp (no matter how large was $\gamma_0$).
Thus, we obtain for $\gamma_1\gg\gamma_c$ the result $\bar\sigma^{(2)}_{\gamma_0,\gamma_1}(\infty)
\to\bar\sigma^{(1)}_{\gamma_1}(\infty)$. Only if the second step is small can there be any positive residual stress from the first step. 
The BKZ limit of Eq.~\eqref{dt0-ramp} is only reached if both
$\Delta t,\delta t\ll\tau_0$: even for arbirarily high shear rate,
Eq.~\eqref{mz} implies that $\tau_0$ governs the fastest relaxation
possible for $\phi(t,t')$. If both the time between the ramps and the
duration of the second ramp are short compared to this relaxation, some
memory can be kept of the first step.

\begin{figure}
\includegraphics[width=.9\linewidth]{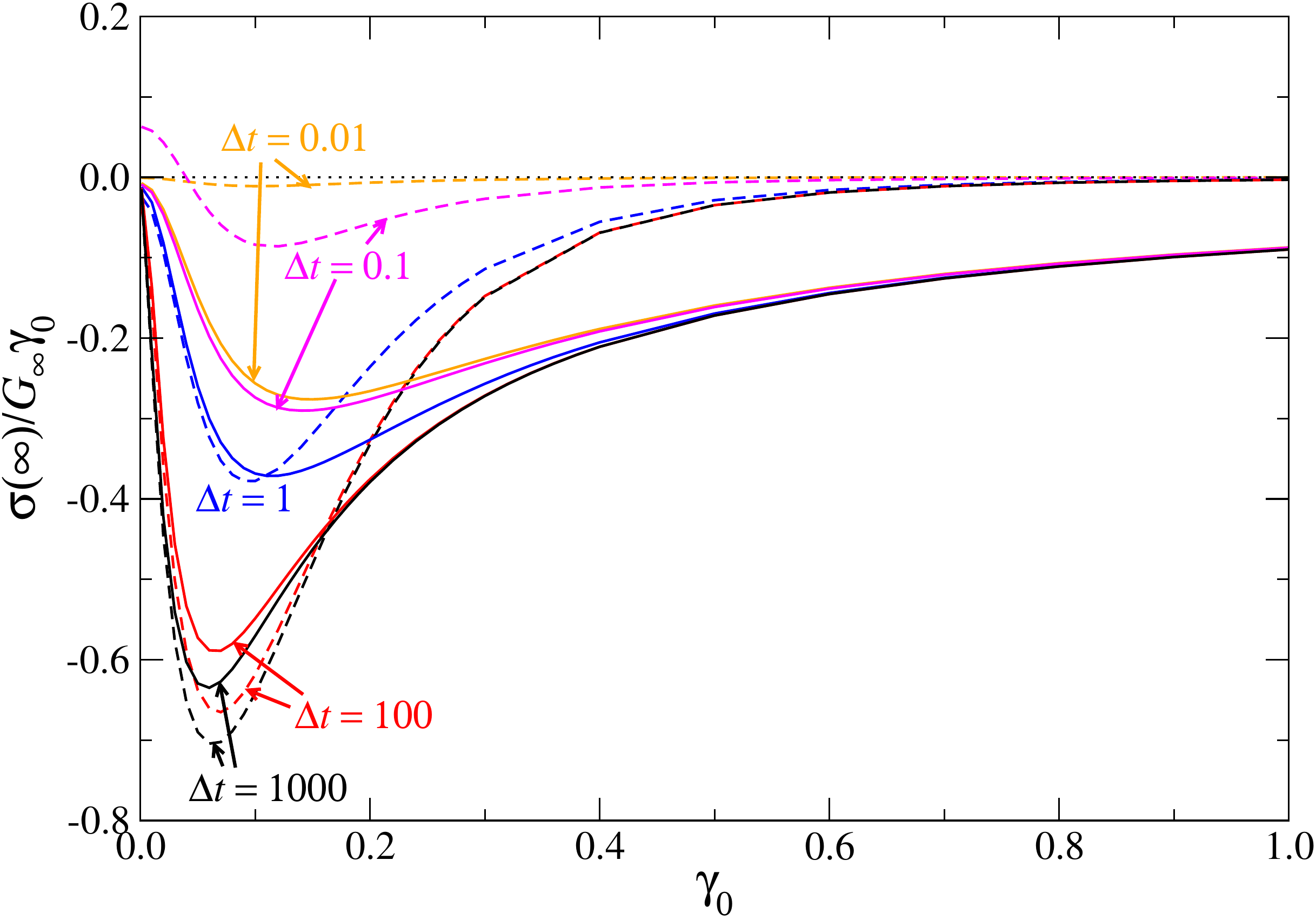}
\caption{\label{recov}
  Stress recovery after exact step strain reversal: $\bar\sigma(\infty)$
  as a function of and normalized by the elastic response $G_\infty\gamma_0$ for
  waiting times $\Delta t=0.01$, $0.1$, $1$, $100$, and $1000$
  (top to bottom). F12 model, $\varepsilon=10^{-4}$.
  Solid lines:
  ramp model with $\delta t=1$.
  Dashed lines: $\sigma(\infty)$ for $\delta$-step model.
}
\end{figure}

We next specialize to the case of exactly canceling double
step strains, $\gamma_1=-\gamma_0$.
In Fig.~\ref{recov} we show the residual stress normalized by the
linear elastic stress, $\bar\sigma(\infty)/(G_\infty\gamma_0)$
as a function of $\gamma_0$, using a ramp time $\delta t=1$ chosen such that
the bare Peclet number is modest ($\le1$) within the
parameter range addressed. 
The results contrast strongly the class of constitutive equations based on the BKZ
form Eq.~\eqref{bkz}, for which all the curves in Fig.~\ref{recov} would be
identically zero.
In the linear response regime, $\gamma_0\to0$, the ratio $\bar\sigma(\infty)/\gamma_0$ does not vanish but attains a limiting value that depends on $\Delta t$. Around $\gamma_0\approx\gamma_c$, a minimum occurs that becomes
more pronounced if $\Delta t$ is increased, until it saturates for time delays beyond the relaxation time of the single step-strain response. At large strains $\gamma_0\gtrsim\gamma_c$, $\bar\sigma(\infty)$ 
approaches the same saturating value as following a single step, which shows up here as a slowly decaying  $1/\gamma_0$ asymptote at large strains that is independent of $\Delta t$. (The delta-step results, shown in Fig.~\ref{recov} as dashed lines, will be discussed later.) 

\begin{figure}
\includegraphics[width=.9\linewidth]{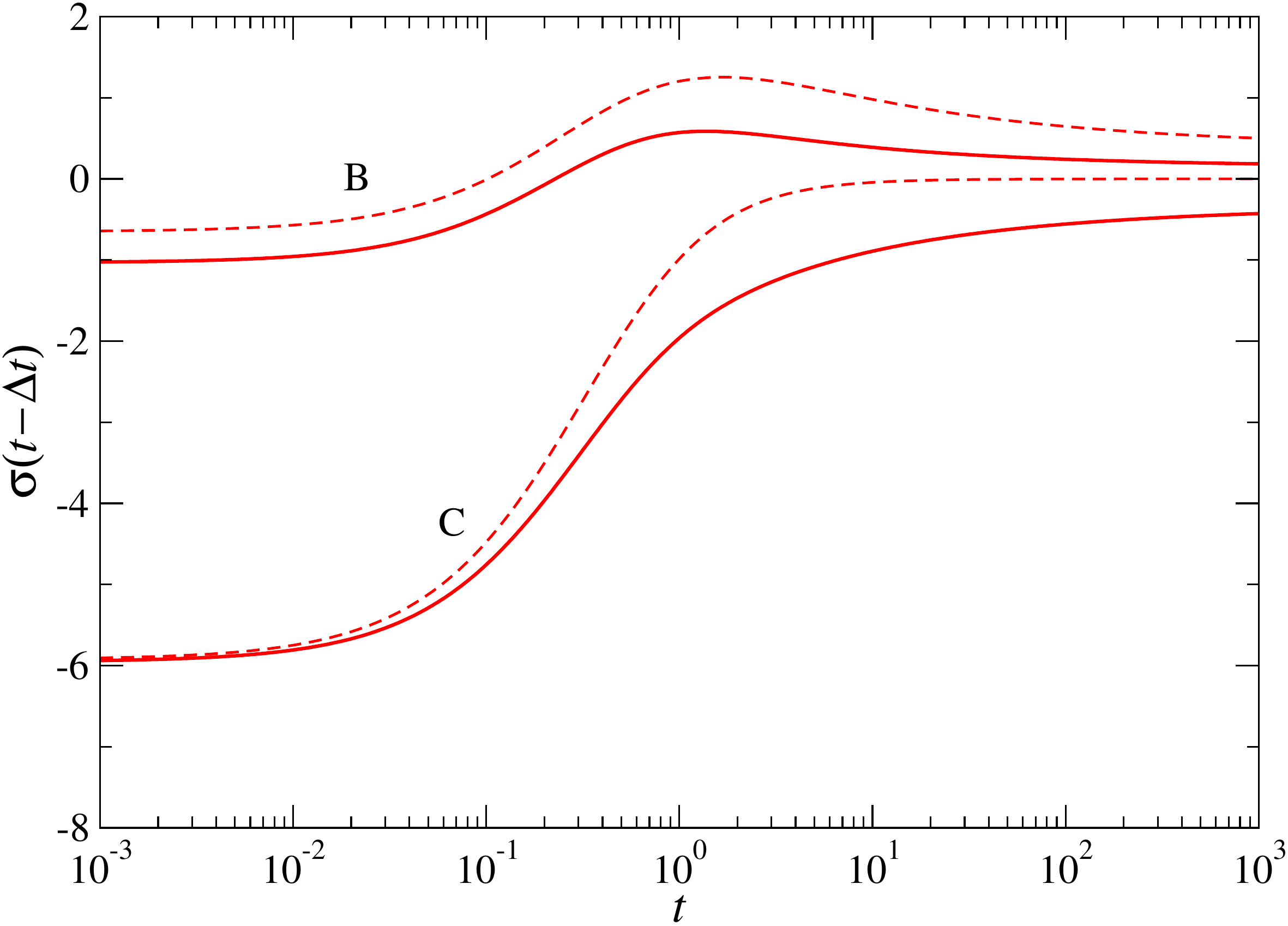}
\caption{\label{dblstep-kbkz-ramp}
  Time-dependent stress relaxation $\sigma(t)$ after a double strain ramp,
  with $\delta t=0.01$, F12 model for
  $\varepsilon=10^{-4}$. The first strain has a magnitude $\gamma_0=0.1$,
  and the second $\gamma_1=-\gamma_0/2$ (label B), respectively
  $\gamma_1=-\gamma_0$ (label C). Solid lines: MCT predictions.
  Dashed lines: BKZ prediction, using MCT's single-strain ramp result
  as an input to Eq.~\eqref{bkz}.
}
\end{figure}

Figure~\ref{dblstep-kbkz-ramp} shows the time-dependent stress relaxation
after reversed step strains, for the two cases $\gamma_1=-\gamma_0/2$
(sometimes classified as type~B in the rheology literature \cite{Venerus1990,Wagner1998}), and $\gamma_1=-\gamma_0$ (type~C).
A strain ramp with $\delta t=0.01$ was chosen, and an initial strain of
$\gamma_0=0.1$. We note that $\bar\sigma(t)$ can be a non-monotonous function,
seen here for the type~B curve which exhbits a maximum at $t\approx1$.
Such nonmonotonic variation arises when the relaxations following the
two steps have different time scales.
According to the discussion of Fig.~\ref{dstepx-ramp}, such an interference
of the two strain ramps is only possible
if $\Delta t$ and $\delta t$ are comparable to $\tau_0$.

Dashed lines in Fig.~\ref{dblstep-kbkz-ramp} show the BKZ prediction
according to Eq.~\eqref{bkz}.
The BKZ factorization overestimates $\sigma(t)$ in type~B and similar partially
reversed double-step strains. This is a known effect in polymer melt
rheology \cite{LarsonValesano,WagnerEhrecke}, where qualitatively similar
results have been found and the deviation has been attributed to
visco-\emph{anelastic} effects \cite{LarsonValesano} or irreversible
breaking of networks (i.e., a persistent loss of memory)
\cite{WagnerStephenson}. In our model, the difference arises because the
system \emph{keeps} memory, but an analysis is not so simple due to the
complicated structure of our constitutive equation.
The deviations from BKZ are not particularly marked for small and intermediate
times. At large times, the BKZ approximation for type~C differs qualitatively
from our model, as explained in connection with Fig.~\ref{recov}. Note that
in a liquid, this difference can only be seen if the quiescent relaxation
time is large enough.

\section{Step Strains as Discontinuities}

We now discuss how the ramp-step model discussed so far differs
from a treatment of instantaneous step strains as delta functions
within the schematic MCT approach.

For a double step strain, one can formally set $\dot\gamma(t)=\gamma_0\delta(t)
+\gamma_1\delta(t-\Delta t)$. The total strain at $t>\Delta t$ is hence
$\gamma=\gamma_0+\gamma_1$. In formulating the equations for this
idealization, there appear damping functions of the form $h_{t0}=h[\gamma_{t0}]$
with
\begin{equation}\label{delta}
  \gamma_{t0}=\gamma_0\int_0^t\delta(\tau)\,d\tau+\gamma_1
  \quad\text{for $t>\Delta t$}
\end{equation}
The integral in this equation can take on any value between zero and unity, depending
on the chosen sequence of functions used to construct the 
distribution $\delta(\tau)$ in the limit $\delta t\to0$. 
The fact that Eq.~\eqref{delta} is formally undefined is of course itself a warning that step strains in MCT are not what they seem.

One obvious choice is to
include the full strain in the integral, hence $\gamma_{t0}=\gamma_0+\gamma_1$.
Another common interpretation is to set the integral to $1/2$
(interpreting it as a unit step function whose value at the jump discontinuity
is set to the midpoint in reference to Dirichlet's theorem of Fourier
representations). We will make the latter choice in what follows (but mention
the differences where these are marked). The midpoint interpretation was already used in
previous MCT calculations for single step strains \cite{Brader2007,Brader2009}, although there the factor $1/2$ can anyway be
absorbed by a redefinition of $\gamma_c$. 

In Eq.~\eqref{gk} our choice leads to
\begin{equation}
  \sigma^{(2)}_{\gamma_0,\gamma_1}(t)=
  \gamma_0G^{(2)}_{\gamma_0,\gamma_1}(t)+
  \gamma_1G^{(1)}_{\gamma_1}(t-\Delta t)
\end{equation}
valid for $t>\Delta t$. The second term is the response to a single delta-step
of magnitude $\gamma_1$, as defined previously via Eq.~\eqref{GDEF} in terms of the correlator $\phi^{(1)}_{\gamma_0}(t)$ that obeys Eq.~\eqref{phi1}.
The first term $G^{(2)}(t)$ cannot be expressed in terms of single-step
response functions alone and therefore, as with the ramp-step approach, we are dealing with a constitutive model that is not of BKZ factorizable form. The corresponding correlator $\phi^{(2)}_{\gamma_0,\gamma_1}(t)$
obeys Eq.~\eqref{eomphi30} evaluated at $x=1/2$ (see Appendix).

Before describing the numerical results based on these equations we consider first the limit
$\Delta t\to0$. Then, the first integral in Eq.~\eqref{eomphi30} can be
neglected, and that equation attains the form of Eq.~\eqref{phi1}.
We get
\begin{equation}\label{dt0-step}
  \lim_{\Delta t\to0}\sigma^{(2)}_{\gamma_0,\gamma_1}(t)
  =\gamma_0G^{(1)}_{\gamma_0+2\gamma_1}(t)
  +\gamma_1G^{(1)}_{\gamma_1}(t-\Delta t)
\end{equation}
Note that this is not equivalent to a single step of the combined magnitude.
In contrast to the BKZ form, Eq.~\eqref{bkz},
the first term is not obviously split into two contributions of the
form $\sigma^{(1)}(t)$. Note also how the combination $\gamma_0/2+\gamma_1$ enters
in the first term. This is a direct result of our midpoint treatment of the integral in Eq.~\ref{delta}.
With this choice only, it is easily established that for an exactly reversed step strain ($\gamma_1=-\gamma_0$), as $\Delta t\to0$, the stress response vanishes identically for all strain amplitudes. (Interpreting the integral of
Eq.~\eqref{delta} instead to yield $\gamma_0+\gamma_1$, this cancellation would only occur inside the linear-response regime.)
Should this be considered a desirable property -- which it certainly would be in a model where the immediate response to a step strain involved no plastic deformation -- then this is a good reason to choose the midpoint treatment of the integral.

\subsection{Results for Delta-Step Approach}

For a single delta-step strain, the correlator obeys Eq.~\eqref{phi1} introduced previously. From this it is straightforward to establish that the residual nonergodicity parameter, long after such a step strain, obeys
\begin{equation}\label{siginf-step}
  f^{(1)}_{\gamma_0}=\lim_{t\to\infty}\phi^{(1)}_{\gamma_0}(t)
  =\frac{h(\gamma_0/2)m_0[f_0]}{1+h(\gamma_0/2)m_0[f_0]}
\end{equation}
From this follows the ultimate residual stress
\begin{equation}
\sigma^{(1)}(t\to\infty)\to\sigma(\infty)=\gamma_0v_\sigma
(f^{(1)}_{\gamma_0})^2
\end{equation}
which differs markedly in functional form from the ramp-step case, Eq.~\ref{siginf-ramp}. Instead of saturating at large strains, as there, the residual stress now has a maximum at $\gamma_0\approx\gamma_c$ and falls to zero at large strain amplitudes, as was found previously in both ITT-MCT and schematic approaches \cite{Brader2007,Brader2009,Brader2009note}. 
(This result is also reminiscent of the Doi-Edwards theory
and experiments on entangled polymer melts \cite{Wang2006}.)
The prediction is shown in Fig.~\ref{stepstrain-dtsiginf} as a dashed blue line
alongside data from the ramp-step approach.

Through the fact that $h$ decreases indefinitely for large strains,
the delta-step model incorporates an extreme form of memory loss. This
leads to plastic deformation in the time following the step strain that
is very effective in reducing the stress.
This contrasts with the ramp-step case (which, as explained above, is
closely related to the switchoff of steady shear): there, the memory loss
during the application of the strain is self-limited due to the plastic flow
already occuring in that time interval.

\begin{figure}
\includegraphics[width=.9\linewidth]{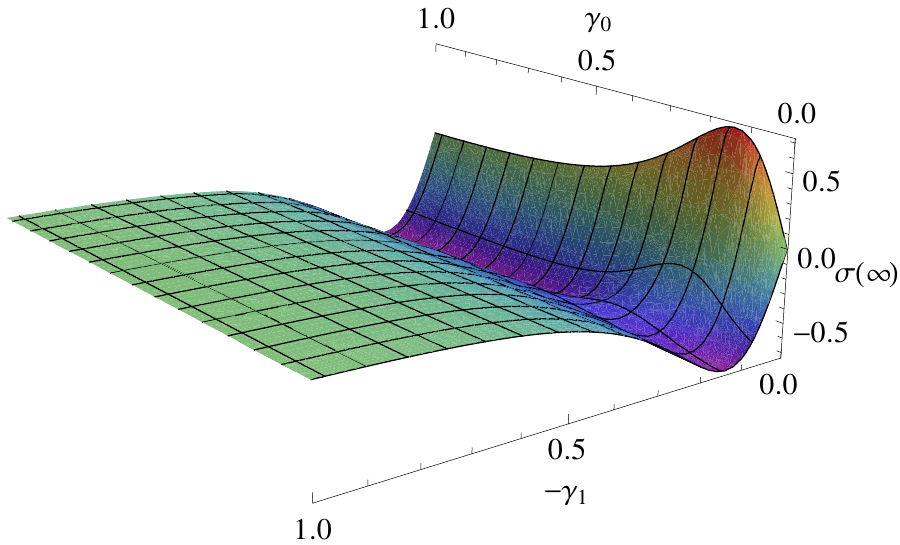}\\
\includegraphics[width=.9\linewidth]{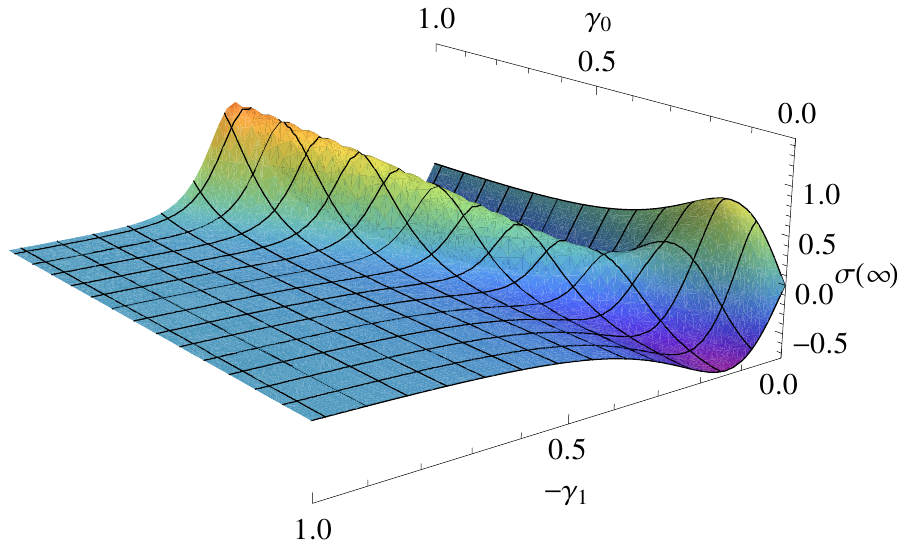}
\caption{\label{dstepx}
  Infinite-time stress plateau $\sigma(\infty)$ after reversing double delta-step
  strains in the glass (F12 model, $\varepsilon=10^{-4}$),
  as a function of the first and second
  step sizes, $\gamma_0>0$ and $|\gamma_1|=-\gamma_1$.
  Time between the strains: $\Delta t=10$ (top graph) and $\Delta t=1$ (bottom).
}
\end{figure}

Turning now to double step strain, the residual stress $\sigma^{(2)}(\infty)$ as a function of the two strain amplitudes is shown for reversing step strains in Fig.~\ref{dstepx},
where $\gamma_0>0$ but $\gamma_1<0$ in analogy to the case discussed in
Fig.~\ref{dstepx-ramp} for the ramp model. The appearance of a maximum in
the single-step response (dashed curve in Fig.~\ref{stepstrain-dtsiginf}),
reappearing here as the boundaries of the surface at $\gamma_0=0$ or
$\gamma_1=0$, imposes a nonmonotonic variation of $\sigma^{(2)}(\infty)$
also when both steps are finite.
As noted in connection with Eq.~\eqref{dt0-step}, along the line
$\gamma_0=-\gamma_1$, $\sigma(\infty)\approx0$ is expected for
small $\Delta t$.

Along the line $\gamma_0=-2\gamma_1$, a pronounced
maximum occurs for small enough $\Delta t$ that arises because of a
cancellation of the $\gamma_0/2+\gamma_1$ factor in the damping function
of Eq.~\eqref{dt0-step}. Interpreting the one-sided $\delta$-integral
as unity would shift this maximum to the line $\gamma_0=-\gamma_1$.
No such maximum was found for the ramp calculations. It arises from
a cancellation of strain-induced damping terms entering the memory functions
and can hence be interpreted as an ``echo'' effect: two instantaneous
anelastic deformations can cancel if applied with a time lag
$\Delta t\lesssim\tau_0$ such that no intermediate Brownian motion destroys
the reversibility. In the ramp model, on the contrary,
plastic deformation is already present during the application of the
strain steps, and these are not reversible.

Just as in the ramp-step case (Fig.~\ref{dstepx-ramp}), the plots in Fig.~\ref{dstepx} show a strong departure from the form expected with BKZ-type constitutive equations. (Recall that in the latter,
Eq.~\eqref{bkz} evaluates to
$\sigma^{(2)}(\infty)=\sigma^{(1)}_{\gamma_0+\gamma_1}(\infty)$, so that
the graph in Fig.~\ref{dstepx} should be inflection-symmetric around
the line $\gamma_0=-\gamma_1$.)
However the form of this departure is strongly different from that found in the ramp-step approach. For instance if $\Delta t$ is large enough such that the response to the initial stress
has decayed to its long-time plateau $\sigma^{(1)}_{\gamma_0}(\infty)$,
the second strain superimposes a negative contribution such that the
total residual stress can become negative even if $\gamma_0+\gamma_1>0$.
But instead of a large second strain erasing all memory of the first (larger)
step, the long-time asymptotes of the stresses essentially add:
$\lim_{\Delta t\to\infty}\sigma^{(2)}_{\gamma_0,\gamma_1}(\infty)
=\sigma^{(1)}_{\gamma_0}(\infty)+\sigma^{(1)}_{\gamma_1}(\infty)$,
i.e., the delta-step model keeps a persistent memory of the residual stress of
the first step.

For the case of exact strain reversal, results for the residual stress using the delta-step approach are shown alongside those for ramp-steps in Fig.~\ref{recov}. Apart from the much faster decay at large amplitudes (which is the counterpart of the very rapidly decreasing residual stress after a single step, discussed above) the two methods give broadly similar results.

\section{Conclusions}

For many strongly viscoelastic materials, such as polymers, the step-strain response is a cornerstone of nonlinear rheology. This holds both experimentally, where single and double step strains are major diagnostic tests, and in the development of theories. For instance, theories of BKZ type create an entire constitutive model solely from the nonlinear response to a single step strain, while double step strains have long been advocated specifically to probe any breakdown of BKZ-like superposition precepts (see, e.g. Ref.~\citep{DoiEdwards}). Across the whole of polymer viscoelasticity theory, the assumption that step-strain responses can be measured experimentally, to sufficient accuracy to inform these theories, goes largely unquestioned.

In this paper we have presented a careful analysis of both single and double step strains within the schematic MCT approach to the rheology of colloidal glasses. These are yield-stress materials, whose response to flow includes a strong contribution from plasticity: structural rearrangements are caused by straining the material. As discussed in the introduction, and amply confirmed by the calculations presented above, BKZ precepts can fail drastically in the description of such plastic materials. 
(This was also noted for a mesoscopic model of soft glasses in Ref.~\citep{Sollich1998}.)
We have discussed this in terms of the residual stress $\sigma(\infty)$
remaining in a glass that has been subject to an exactly reversed double
step strain. BKZ models predicts this residual stress to vanish. Within
MCT, a finite residual stress remains: it is non-vanishing because
plastic flow has taken place in the time between the strains, something
not accounted for in the BKZ constitutive equation.
This may have interesting consequences for other types of deformation
like oscillatory shear.

Perhaps more surprisingly, our work exposes the fact that in materials where plasticity is important, the very concept of a step strain is problematic. Specifically we have shown that within the mathematics of our schematic MCT model, the response to a discontinuous strain is different from the response to a finitely ramped one, even if the limit is then taken of the ramp rate tending to infinity. This is because in strong flows the relaxation time of the material is inversely proportional to strain rate. Accordingly, plastic deformation does not vanish in the fast ramp limit, whereas within our chosen equations it is (apparently) absent during the imposition of a mathematically discontinuous step.

A possible interpretation of the different representations of strain steps
in schematic MCT is this: in the delta-function model, the strain in the
system just after a step is primarily a large nonlinear elastic (anelastic)
strain that then converts to plastic strain as memory loss kicks in.
(Of course, this assumption is questionable whenever $\gamma\gg\gamma_c$,
at least for hard-core potentials. It may be of relevance for soft
particles.)
The corresponding strain energy is relaxed via subsequent plastic
deformations, as expressed through a strain-induced decay of correlation
functions. The conversion of anelastic to deformation to plastic
rearrangements is particularly effective at large strains, so that
most of the stresses relax, and $\sigma(\infty)$ is almost zero
(dashed lines in Fig.~\ref{stepstrain-dtsiginf}).
The strain-ramp model, on the other hand, assumes that
plastic deformations are predominant already during the application of
the strains. Consequently, the system is always fluidized in these
time intervals, to the extent of completely erasing memory of any past
flow history. While the delta-step model captures the fact that
anelastic deformations can partially be transformed into plastic ones,
but partially be recovered in a double-step strain setup, the ramp model
captures that any strain much larger than $\gamma_c$ will lead to a
full loss of reversibility. As a consequence, stresses can no longer
be relaxed as effectively, and a saturation of the
$\bar\sigma(\infty)$ curves is seen
in Fig.~\ref{stepstrain-dtsiginf} for $\gamma_0\gtrsim\gamma_c$.

Here, it should be noted, that the ramp model excludes some of the
anelastic deformation mechanisms present in
the full MCT. There the analog to Eq.~\eqref{GDEF} quite naturally
leads to a vertex $v_\sigma(\gamma_{tt'})$ that itself depends on strain
through a wave-vector advection mechanism \cite{Brader2008,Brader2009}. Furthermore, this vertex shows
negative contributions for strains $\gamma\approx\gamma_c$, manifesting
themselves in the appearance of a stress overshoot in startup flow
\cite{Zausch2008}. Schematic
models that include such anelastic contributions can be devised
and will suppress the residual
stress $\bar\sigma^{(1)}(\infty)$ shown in Fig.~\ref{stepstrain-dtsiginf}
for large $\gamma_0$, possibly reinstating a maximum around $\gamma_c$.

Although we cannot be certain, the mathematical issue discussed above might reflect a genuine physical one, of whether the bare Peclet number, $\dot\gamma\tau_0\ll1$, is large or small during the rapid straining event that any experimental `step' represents. 
(Here $\tau_0$ is a non-glassy, microscopic relaxation time and $\dot\gamma$ the shear rate.) In most `step strain' experiments the bare Peclet number is unlikely to be large (see e.g.\ Ref.~\cite{Pham2008}); hence if this reasoning is correct, the results we have presented for finite ramp rates, rather than those for delta-function steps, are more pertinent. To confirm this point, it would be very interesting to see more experiments on step strain, particularly double step strain, in colloidal systems.
As discussed above, the dependence of the residual stress $\sigma^{(2)}(\infty)$
after a double step strain on the individual strains $\gamma_0$ and $\gamma_1$
can provide a sensitive test for the different scenarios.

One has to keep in mind that real glassy materials will be subject to aging.
In our discussion we implied that the initial unstrained configuration was
equilibrated (i.e., representative of the Boltzmann ensemble, even if the dynamics is
nonergodic in the glass). The usual rheology protocol (pre-shear to fluidize
the system, followed by a certain waiting time $t_w$) does not ensure this.
We expect however, that for sufficiently long $t_w$ the schematic MCT results
can at least qualitatively or even semi-quantitatively be compared to
experiment, as was previously found true for flow curves in steady state \cite{Crassous2006}. Note that the imposition of a flow or step-strain displaces the system from Boltzmann equilibrium even if it was previously equilibrated as MCT assumes. The subsequent relaxation (e.g. between two steps in double step strain) may also have aging-like features, which MCT should capture.





We thank R.~G.~Larson for valuable discussions during early stages of this work.
This work was partially supported by Deutsche Forschungsgemeinschaft,
FOR1394 project P3.
ThV thanks for funding from
Helmholtz-Gemeinschaft (HGF Hochschul-Nachwuchsgruppe VH-NG~406),
and Zukunftskolleg der Universit\"at Konstanz.
MEC holds a Royal Society Research Professorship and is funded via
EPSRC EP/E030173.

\begin{appendix}
\section{Appendix: Numerical Scheme}

For a strain history comprised of piecewise constant shear rates,
one is led to split the integrals in Eqs.~\eqref{gk} and \eqref{mz} at the
step discontinuities in $\dot\gamma(t)$. Specifically, set
$\dot\gamma(t)=\dot\gamma^{(m)}$ for $t_m<t<t_{m+1}$ to obtain
\begin{equation}
  \sigma(t)=\sum_{m=0}^{n-1}\dot\gamma^{(m)}
  \int_{t_m}^{t_{m+1}}G^{(nm)}(t,t')\,dt'
  +\dot\gamma^{(n)}\int_{t_n}^tG^{(nn)}(t,t')\,dt'
\end{equation}
for $t_n<t<t_{n+1}$.
We have assumed $\dot\gamma(t)=0$ for $t<t_0$ and have introduced
functions $G^{(mn)}(t,t')$ that are given by
Eq.~\eqref{GDEF} with
transient correlation functions $\phi^{(mn)}(t,t')$ defined for
$t_m<t<t_{m+1}$, $t_n<t'<t_{n+1}$. For these, the equations of motion read
\begin{multline}\label{phimn}
  \tau_0\partial_t\phi^{(mn)}(t,t')+\phi^{(mn)}(t,t')
  \\
  +h_{tt'}\int_{t'}^{t_{m+1}}h_{tt''}m[\phi^{(mn)}(t,t'')]
  \partial_{t''}\phi^{(nn)}(t''-t')\,dt''
  \\
  +h_{tt'}\sum_{l=m+1}^{n-1}\int_{t_l}^{t_{l+1}}h_{tt''}
  m[\phi^{(ml)}(t,t'')]\partial_{t''}\phi^{(ln)}(t'',t')\,dt''
  \\
  +h_{tt'}\int_{t_n}^th_{tt''}m[\phi^{(mm)}(t-t'')]
  \partial_{t''}\phi^{(mn)}(t'',t')\,dt''=0\,.
\end{multline}
Note that the correlation functions are transient ones, and as such
are only affected by the strains applied between the times of their
arguments. Hence, the $\phi^{(mm)}(t,t')\equiv\phi^{(mm)}(t-t')$ are
steady-state quantities (quiescent or steady-shear) that depend only on the
time difference. Equations \eqref{phimn} are complemented by initial
conditions obtained from the fact that correlation functions are continuous
in both time arguments.

\begin{figure}
\includegraphics[width=.9\linewidth]{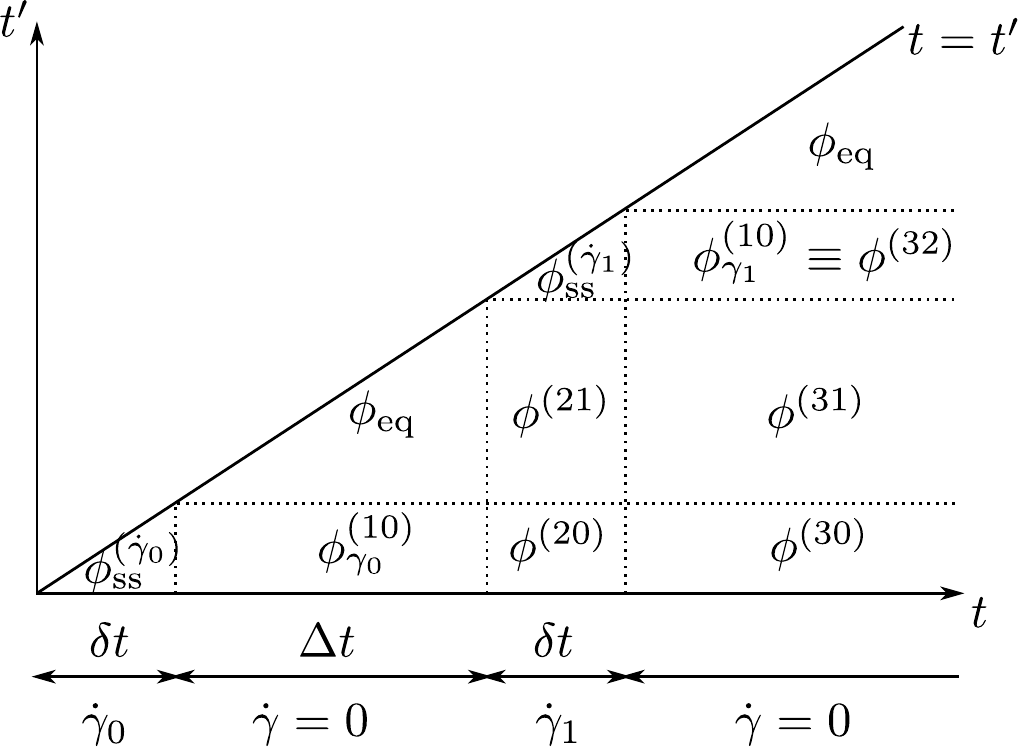}
\caption{\label{schema}
  Schematic view of the $(t,t')$ plane indicating the various transient
  correlation functions appearing in the MCT equations for a double strain
  ramp. See text for details.
}
\end{figure}

We now specialize
Eq.~\eqref{phimn} to one or two strain ramps, $\dot\gamma^{(0)}=
\dot\gamma_0$, $\dot\gamma^{(2)}=\dot\gamma_1$, and all other
$\dot\gamma^{(m)}=0$. Then we are left with nine different
correlation functions, three of which are steady-state ones:
$\phi^{(00)}\equiv\phi_\text{ss}^{(\dot\gamma_0)}$,
$\phi^{(22)}\equiv\phi_\text{ss}^{(\dot\gamma_1)}$,
and $\phi^{(11)}\equiv\phi^{(33)}\equiv\phi_0$. Figure~\ref{schema}
gives a schematic overview. It is straightforward to specialize
Eq.~\eqref{phimn} to the cases needed in evaluating $\bar\sigma^{(1)}(t)$
and $\bar\sigma^{(2)}(t)$, the single- and double-ramp response,
\begin{align}
  \bar\sigma^{(1)}(t)&=\dot\gamma_0\int_0^{\delta t}G^{(10)}_{\gamma_0}(t,t')
  \,dt'
  &&\text{for $t>\delta t$,}\\
  \bar\sigma^{(2)}(t)&=\dot\gamma_0\int_0^{\delta t}
  G^{(30)}_{\gamma_0,\gamma_1}(t,t')\,dt'
  \\ \nonumber
  &+\dot\gamma_1\int_{\Delta t+\delta t}^{\Delta t+2\delta t}
  G^{(32)}_{\gamma_1}(t,t')\,dt'
  &&\text{for $t>\Delta t+2\delta t$.}
\end{align}
Here we have introduced interval
lengths $\delta t$ and $\Delta t$ as indicated in Fig.~\ref{schema}.
We can identify $\phi^{(32)}(t,t')$ with the suitably shifted
and extended
$\phi^{(10)}(t-\Delta t-\delta t,t'-\Delta t-\delta t)$ evaluated
for $\dot\gamma_1$ instead of $\dot\gamma_0$. The function
$\phi^{(10)}(t,t')$ defines the response to a single strain ramp,
while $\phi^{(30)}(t,t')$ includes cross-correlations between the
two ramps.

If one considers the limit $\delta t\to0$, the equations of motion
for $\phi^{(10)}$ and $\phi^{(30)}$ simplify as some of the integrals
appearing in Eq.~\eqref{phimn} vanish. Introducing $x=t'/\delta t$
and $\bar\phi(t,x)=\bar\phi(t,t')$ as in the main text, we get
for the single-ramp response
\begin{multline}\label{eomphi10}
  [1+\tau_0\partial_t]\bar\phi^{(10)}_{\gamma_0}(t,x)
  \\
  +h(\gamma_0(1-x))\int_0^tm[\phi_0(t-t'')]\partial_{t''}
  \bar\phi^{(10)}_{\gamma_0}(t'',x)=0\,,
\end{multline}
a one-parameter family of functions for $x\in[0,1]$.
Similarly,
\begin{multline}\label{eomphi30}
  [1+\tau_0\partial_t]\bar\phi^{(30)}_{\gamma_0,\gamma_1}(t,x)
  \\
  +h(\gamma_1+\gamma_0(1-x))h(\gamma_1)
  \int_0^{\Delta t}m[\phi^{(31)}_{\gamma_0,\gamma_1}(t,t'')]
  \partial_{t''}\bar\phi^{(10)}_{\gamma_0}(t'',x)\,dt''
  \\
  +h(\gamma_1+\gamma_0(1-x))\int_{\Delta t}^t m[\phi_0(t-t'')]
  \partial_{t''}\bar\phi^{(30)}_{\gamma_0,\gamma_1}(t'',x)\,dt''=0
\end{multline}
There appears a cross-correlator in this equation, whose equation (in
the limit $\delta t\to0$) reads
\begin{multline}\label{eomphi31}
  [1+\tau_0\partial_t]\phi^{(31)}(t,t')
  \\
  +h(\gamma_1)^2\int_{t'}^{\Delta t}m[\phi^{(31)}(t,t'')]
  \partial_{t''}\phi_0(t''-t')\,dt''
  \\
  +h(\gamma_1)\int_{\Delta t}^tm[\phi_0(t-t'')]
  \partial_{t''}\phi^{(31)}(t'',t')\,dt''=0\,.
\end{multline}
This cross-correlator plays an important role in establishing the
non-trivial dependence on $\Delta t$ of the MCT double-ramp results.
It is a result of these and similar cross-correlations
that the MCT constitutive equation is not of BKZ type.

Deriving the equivalent set of equations for the case of delta-function
steps, it is easy to see that again Eqs.~\eqref{eomphi10} to \eqref{eomphi31}
appear, but evaluated only at one specific value of $x$; in the notation
of the main text $\phi^{(1)}(t)\equiv\bar\phi^{(10)}(t,1/2)$ and
$\phi^{(2)}(t)\equiv\bar\phi^{(30)}(t,1/2)$. The cross-correlator
$\phi^{(31)}(t,t')$ appears identically in both models.

\end{appendix}

\footnotesize{
\bibliography{rsc} 

\providecommand*{\mcitethebibliography}{\thebibliography}
\csname @ifundefined\endcsname{endmcitethebibliography}
{\let\endmcitethebibliography\endthebibliography}{}
\begin{mcitethebibliography}{42}
\providecommand*{\natexlab}[1]{#1}
\providecommand*{\mciteSetBstSublistMode}[1]{}
\providecommand*{\mciteSetBstMaxWidthForm}[2]{}
\providecommand*{\mciteBstWouldAddEndPuncttrue}
  {\def\EndOfBibitem{\unskip.}}
\providecommand*{\mciteBstWouldAddEndPunctfalse}
  {\let\EndOfBibitem\relax}
\providecommand*{\mciteSetBstMidEndSepPunct}[3]{}
\providecommand*{\mciteSetBstSublistLabelBeginEnd}[3]{}
\providecommand*{\EndOfBibitem}{}
\mciteSetBstSublistMode{f}
\mciteSetBstMaxWidthForm{subitem}
{(\emph{\alph{mcitesubitemcount}})}
\mciteSetBstSublistLabelBeginEnd{\mcitemaxwidthsubitemform\space}
{\relax}{\relax}

\bibitem[Larson(1988)]{LarsonCE}
R.~G. Larson, \emph{Constitutive Equations for Polymer Melts and Solutions},
  Butterworths, London, 1988\relax
\mciteBstWouldAddEndPuncttrue
\mciteSetBstMidEndSepPunct{\mcitedefaultmidpunct}
{\mcitedefaultendpunct}{\mcitedefaultseppunct}\relax
\EndOfBibitem
\bibitem[Petekidis \emph{et~al.}(2002)Petekidis, Moussa{\"\i}d, and
  Pusey]{Petekidis2002}
G.~Petekidis, A.~Moussa{\"\i}d and P.~N. Pusey, \emph{Phys. Rev. E}, 2002,
  \textbf{66}, 051402\relax
\mciteBstWouldAddEndPuncttrue
\mciteSetBstMidEndSepPunct{\mcitedefaultmidpunct}
{\mcitedefaultendpunct}{\mcitedefaultseppunct}\relax
\EndOfBibitem
\bibitem[Haw(2004)]{Haw2004}
M.~D. Haw, \emph{Phys. Rev. Lett.}, 2004, \textbf{92}, 185506\relax
\mciteBstWouldAddEndPuncttrue
\mciteSetBstMidEndSepPunct{\mcitedefaultmidpunct}
{\mcitedefaultendpunct}{\mcitedefaultseppunct}\relax
\EndOfBibitem
\bibitem[Crassous \emph{et~al.}(2006)Crassous, Siebenb{\"u}rger, Ballauff,
  Drechsler, Henrich, and Fuchs]{Crassous2006}
J.~J. Crassous, M.~Siebenb{\"u}rger, M.~Ballauff, M.~Drechsler, O.~Henrich and
  M.~Fuchs, \emph{J. Chem. Phys.}, 2006, \textbf{125}, 204906\relax
\mciteBstWouldAddEndPuncttrue
\mciteSetBstMidEndSepPunct{\mcitedefaultmidpunct}
{\mcitedefaultendpunct}{\mcitedefaultseppunct}\relax
\EndOfBibitem
\bibitem[Pham \emph{et~al.}(2006)Pham, Petekidis, Vlassopoulos, Egelhaaf,
  Pusey, and Poon]{Pham2006}
K.~N. Pham, G.~Petekidis, D.~Vlassopoulos, S.~U. Egelhaaf, P.~N. Pusey and
  W.~C.~K. Poon, \emph{Europhys. Lett.}, 2006, \textbf{75}, 624--630\relax
\mciteBstWouldAddEndPuncttrue
\mciteSetBstMidEndSepPunct{\mcitedefaultmidpunct}
{\mcitedefaultendpunct}{\mcitedefaultseppunct}\relax
\EndOfBibitem
\bibitem[Besseling \emph{et~al.}(2007)Besseling, Weeks, Schofield, and
  Poon]{Besseling2007}
R.~Besseling, E.~R. Weeks, A.~B. Schofield and W.~C.~K. Poon, \emph{Phys. Rev.
  Lett.}, 2007, \textbf{99}, 028301\relax
\mciteBstWouldAddEndPuncttrue
\mciteSetBstMidEndSepPunct{\mcitedefaultmidpunct}
{\mcitedefaultendpunct}{\mcitedefaultseppunct}\relax
\EndOfBibitem
\bibitem[Isa \emph{et~al.}(2007)Isa, Besseling, and Poon]{Isa2007}
L.~Isa, R.~Besseling and W.~C.~K. Poon, \emph{Phys. Rev. Lett.}, 2007,
  \textbf{98}, 198305\relax
\mciteBstWouldAddEndPuncttrue
\mciteSetBstMidEndSepPunct{\mcitedefaultmidpunct}
{\mcitedefaultendpunct}{\mcitedefaultseppunct}\relax
\EndOfBibitem
\bibitem[Crassous \emph{et~al.}(2008)Crassous, Siebenb{\"u}rger, Ballauff,
  Drechsler, Hajnal, Henrich, and Fuchs]{Crassous2008}
J.~J. Crassous, M.~Siebenb{\"u}rger, M.~Ballauff, M.~Drechsler, D.~Hajnal,
  O.~Henrich and M.~Fuchs, \emph{J. Chem. Phys.}, 2008, \textbf{128},
  204902\relax
\mciteBstWouldAddEndPuncttrue
\mciteSetBstMidEndSepPunct{\mcitedefaultmidpunct}
{\mcitedefaultendpunct}{\mcitedefaultseppunct}\relax
\EndOfBibitem
\bibitem[Pham \emph{et~al.}(2008)Pham, Petekidis, Vlassopoulos, Egelhaaf, Poon,
  and Pusey]{Pham2008}
K.~N. Pham, G.~Petekidis, D.~Vlassopoulos, S.~U. Egelhaaf, W.~C.~K. Poon and
  P.~N. Pusey, \emph{J. Rheol.}, 2008, \textbf{52}, 649--676\relax
\mciteBstWouldAddEndPuncttrue
\mciteSetBstMidEndSepPunct{\mcitedefaultmidpunct}
{\mcitedefaultendpunct}{\mcitedefaultseppunct}\relax
\EndOfBibitem
\bibitem[Zausch \emph{et~al.}(2008)Zausch, Horbach, Laurati, Egelhaaf, Brader,
  Voigtmann, and Fuchs]{Zausch2008}
J.~Zausch, J.~Horbach, M.~Laurati, S.~U. Egelhaaf, J.~M. Brader, {\relax
  Th}.~Voigtmann and M.~Fuchs, \emph{J. Phys.: Condens. Matter}, 2008,
  \textbf{20}, 404210\relax
\mciteBstWouldAddEndPuncttrue
\mciteSetBstMidEndSepPunct{\mcitedefaultmidpunct}
{\mcitedefaultendpunct}{\mcitedefaultseppunct}\relax
\EndOfBibitem
\bibitem[Siebenb{\"u}rger \emph{et~al.}(2009)Siebenb{\"u}rger, Fuchs, Winter,
  and Ballauff]{Siebenbuerger2009}
M.~Siebenb{\"u}rger, M.~Fuchs, H.~Winter and M.~Ballauff, \emph{J. Rheol.},
  2009, \textbf{53}, 707--726\relax
\mciteBstWouldAddEndPuncttrue
\mciteSetBstMidEndSepPunct{\mcitedefaultmidpunct}
{\mcitedefaultendpunct}{\mcitedefaultseppunct}\relax
\EndOfBibitem
\bibitem[{Singh Negi} and Osuji(2009)]{Negi2009}
A.~{Singh Negi} and C.~O. Osuji, \emph{Phys. Rev. E}, 2009, \textbf{80},
  010404(R)\relax
\mciteBstWouldAddEndPuncttrue
\mciteSetBstMidEndSepPunct{\mcitedefaultmidpunct}
{\mcitedefaultendpunct}{\mcitedefaultseppunct}\relax
\EndOfBibitem
\bibitem[Besseling \emph{et~al.}(2010)Besseling, Isa, Ballesta, Petekidis,
  Cates, and Poon]{Besseling2010}
R.~Besseling, L.~Isa, P.~Ballesta, G.~Petekidis, M.~E. Cates and W.~C.~K. Poon,
  \emph{Phys. Rev. Lett.}, 2010, \textbf{105}, 268301\relax
\mciteBstWouldAddEndPuncttrue
\mciteSetBstMidEndSepPunct{\mcitedefaultmidpunct}
{\mcitedefaultendpunct}{\mcitedefaultseppunct}\relax
\EndOfBibitem
\bibitem[Bandyopadhyay \emph{et~al.}(2010)Bandyopadhyay, Mohan, and
  Joshi]{Bandyopadhyay2010}
R.~Bandyopadhyay, P.~H. Mohan and Y.~M. Joshi, \emph{Soft Matter}, 2010,
  \textbf{6}, 1462--1466\relax
\mciteBstWouldAddEndPuncttrue
\mciteSetBstMidEndSepPunct{\mcitedefaultmidpunct}
{\mcitedefaultendpunct}{\mcitedefaultseppunct}\relax
\EndOfBibitem
\bibitem[Negi and Osuji(2010)]{Negi2010}
A.~Negi and C.~Osuji, \emph{EPL}, 2010, \textbf{90}, 28003\relax
\mciteBstWouldAddEndPuncttrue
\mciteSetBstMidEndSepPunct{\mcitedefaultmidpunct}
{\mcitedefaultendpunct}{\mcitedefaultseppunct}\relax
\EndOfBibitem
\bibitem[Laurati \emph{et~al.}(2011)Laurati, Egelhaaf, and
  Petekidis]{Laurati2011}
M.~Laurati, S.~U. Egelhaaf and G.~Petekidis, \emph{J. Rheol.}, 2011,
  \textbf{55}, 673--706\relax
\mciteBstWouldAddEndPuncttrue
\mciteSetBstMidEndSepPunct{\mcitedefaultmidpunct}
{\mcitedefaultendpunct}{\mcitedefaultseppunct}\relax
\EndOfBibitem
\bibitem[Willenbacher \emph{et~al.}(2011)Willenbacher, Vesaratchanon,
  Thorwarth, and Bartsch]{Willenbacher2011}
N.~Willenbacher, J.~S. Vesaratchanon, O.~Thorwarth and E.~Bartsch, \emph{Soft
  Matter}, 2011, \textbf{7}, 5777--5788\relax
\mciteBstWouldAddEndPuncttrue
\mciteSetBstMidEndSepPunct{\mcitedefaultmidpunct}
{\mcitedefaultendpunct}{\mcitedefaultseppunct}\relax
\EndOfBibitem
\bibitem[Derec \emph{et~al.}(2000)Derec, Ajdari, and Lequeux]{Derec2000}
C.~Derec, A.~Ajdari and F.~Lequeux, \emph{Eur. Phys. J. E}, 2000, \textbf{4},
  355--361\relax
\mciteBstWouldAddEndPuncttrue
\mciteSetBstMidEndSepPunct{\mcitedefaultmidpunct}
{\mcitedefaultendpunct}{\mcitedefaultseppunct}\relax
\EndOfBibitem
\bibitem[Coussot \emph{et~al.}(2002)Coussot, Nguyen, Huynh, and
  Bonn]{Coussot2002}
P.~Coussot, Q.~D. Nguyen, H.~T. Huynh and D.~Bonn, \emph{Phys. Rev. Lett.},
  2002, \textbf{88}, 175501\relax
\mciteBstWouldAddEndPuncttrue
\mciteSetBstMidEndSepPunct{\mcitedefaultmidpunct}
{\mcitedefaultendpunct}{\mcitedefaultseppunct}\relax
\EndOfBibitem
\bibitem[Sollich \emph{et~al.}(1997)Sollich, Lequeux, H{\'e}braud, and
  Cates]{SGR}
P.~Sollich, F.~Lequeux, P.~H{\'e}braud and M.~E. Cates, \emph{Phys. Rev.
  Lett.}, 1997, \textbf{78}, 2020--2023\relax
\mciteBstWouldAddEndPuncttrue
\mciteSetBstMidEndSepPunct{\mcitedefaultmidpunct}
{\mcitedefaultendpunct}{\mcitedefaultseppunct}\relax
\EndOfBibitem
\bibitem[Sollich(1998)]{Sollich1998}
P.~Sollich, \emph{Phys. Rev. E}, 1998, \textbf{58}, 738--759\relax
\mciteBstWouldAddEndPuncttrue
\mciteSetBstMidEndSepPunct{\mcitedefaultmidpunct}
{\mcitedefaultendpunct}{\mcitedefaultseppunct}\relax
\EndOfBibitem
\bibitem[Falk and Langer(1998)]{STZ1}
M.~L. Falk and J.~S. Langer, \emph{Phys. Rev. E}, 1998,  7192\relax
\mciteBstWouldAddEndPuncttrue
\mciteSetBstMidEndSepPunct{\mcitedefaultmidpunct}
{\mcitedefaultendpunct}{\mcitedefaultseppunct}\relax
\EndOfBibitem
\bibitem[Falk and Langer(2011)]{STZreview}
M.~L. Falk and J.~S. Langer, \emph{Annu. Rev. Cond. Matt. Phys.}, 2011,
  \textbf{2}, 353--373\relax
\mciteBstWouldAddEndPuncttrue
\mciteSetBstMidEndSepPunct{\mcitedefaultmidpunct}
{\mcitedefaultendpunct}{\mcitedefaultseppunct}\relax
\EndOfBibitem
\bibitem[Fuchs and Cates(2002)]{FuchsCates}
M.~Fuchs and M.~E. Cates, \emph{Phys. Rev. Lett.}, 2002, \textbf{89},
  248304\relax
\mciteBstWouldAddEndPuncttrue
\mciteSetBstMidEndSepPunct{\mcitedefaultmidpunct}
{\mcitedefaultendpunct}{\mcitedefaultseppunct}\relax
\EndOfBibitem
\bibitem[Brader \emph{et~al.}(2007)Brader, Voigtmann, Cates, and
  Fuchs]{Brader2007}
J.~M. Brader, {\relax Th}.~Voigtmann, M.~E. Cates and M.~Fuchs,
  \emph{Phys.~Rev.~Lett.}, 2007, \textbf{98}, 058301\relax
\mciteBstWouldAddEndPuncttrue
\mciteSetBstMidEndSepPunct{\mcitedefaultmidpunct}
{\mcitedefaultendpunct}{\mcitedefaultseppunct}\relax
\EndOfBibitem
\bibitem[Brader \emph{et~al.}(2008)Brader, Cates, and Fuchs]{Brader2008}
J.~M. Brader, M.~E. Cates and M.~Fuchs, \emph{Phys. Rev. Lett.}, 2008,
  \textbf{101}, 138301\relax
\mciteBstWouldAddEndPuncttrue
\mciteSetBstMidEndSepPunct{\mcitedefaultmidpunct}
{\mcitedefaultendpunct}{\mcitedefaultseppunct}\relax
\EndOfBibitem
\bibitem[Brader \emph{et~al.}(2009)Brader, Voigtmann, Fuchs, Larson, and
  Cates]{Brader2009}
J.~M. Brader, {\relax Th}.~Voigtmann, M.~Fuchs, R.~G. Larson and M.~E. Cates,
  \emph{Proc.\ Natl.\ Acad.\ Sci.\ USA}, 2009, \textbf{106}, 15186--15191\relax
\mciteBstWouldAddEndPuncttrue
\mciteSetBstMidEndSepPunct{\mcitedefaultmidpunct}
{\mcitedefaultendpunct}{\mcitedefaultseppunct}\relax
\EndOfBibitem
\bibitem[Bernstein \emph{et~al.}(1963)Bernstein, Kearsley, and Zapas]{BKZ}
B.~Bernstein, E.~A. Kearsley and L.~J. Zapas, \emph{Trans. Soc. Rheol.}, 1963,
  \textbf{7}, 391--410\relax
\mciteBstWouldAddEndPuncttrue
\mciteSetBstMidEndSepPunct{\mcitedefaultmidpunct}
{\mcitedefaultendpunct}{\mcitedefaultseppunct}\relax
\EndOfBibitem
\bibitem[Kaye(1962)]{Kaye}
A.~Kaye, \emph{Non-Newtonian Flow in Incompressible Fluids}, College of
  Aeronautics, Cranfield, CoA note No.~134, 1962\relax
\mciteBstWouldAddEndPuncttrue
\mciteSetBstMidEndSepPunct{\mcitedefaultmidpunct}
{\mcitedefaultendpunct}{\mcitedefaultseppunct}\relax
\EndOfBibitem
\bibitem[Doi and Edwards(1986)]{DoiEdwards}
M.~Doi and S.~F. Edwards, \emph{The Theory of Polymer Dynamics}, Oxford
  University Press, 1986\relax
\mciteBstWouldAddEndPuncttrue
\mciteSetBstMidEndSepPunct{\mcitedefaultmidpunct}
{\mcitedefaultendpunct}{\mcitedefaultseppunct}\relax
\EndOfBibitem
\bibitem[Fuchs and Cates(2009)]{Fuchs2009}
M.~Fuchs and M.~E. Cates, \emph{J. Rheol.}, 2009, \textbf{53}, 957--1000\relax
\mciteBstWouldAddEndPuncttrue
\mciteSetBstMidEndSepPunct{\mcitedefaultmidpunct}
{\mcitedefaultendpunct}{\mcitedefaultseppunct}\relax
\EndOfBibitem
\bibitem[Henrich \emph{et~al.}(2009)Henrich, Weysser, Cates, and
  Fuchs]{Weysser2009}
O.~Henrich, F.~Weysser, M.~E. Cates and M.~Fuchs, \emph{Phil. Trans. R. Soc.
  A}, 2009, \textbf{367}, 5033--5050\relax
\mciteBstWouldAddEndPuncttrue
\mciteSetBstMidEndSepPunct{\mcitedefaultmidpunct}
{\mcitedefaultendpunct}{\mcitedefaultseppunct}\relax
\EndOfBibitem
\bibitem[Kr\"uger \emph{et~al.}(2011)Kr\"uger, Weysser, and Fuchs]{FuchsEPJE}
M.~Kr\"uger, F.~Weysser and M.~Fuchs, \emph{Eur. Phys. J. E}, 2011,
  \textbf{34}, 88\relax
\mciteBstWouldAddEndPuncttrue
\mciteSetBstMidEndSepPunct{\mcitedefaultmidpunct}
{\mcitedefaultendpunct}{\mcitedefaultseppunct}\relax
\EndOfBibitem
\bibitem[fin()]{fingernote}
This is recognized as the specialization of the more general expression in
  Ref.~\protect\cite{Brader2009}, where it was argued that only the invariants
  of the Finger tensor $\boldsymbol B(t,t')$ can enter $h_{tt'}$. For
  deformations that are always simple shear the Finger tensor obeys
  $\text{tr}\boldsymbol B=\text{tr}\boldsymbol B^{-1}=3+\gamma_{tt'}^2$.\relax
\mciteBstWouldAddEndPunctfalse
\mciteSetBstMidEndSepPunct{\mcitedefaultmidpunct}
{}{\mcitedefaultseppunct}\relax
\EndOfBibitem
\bibitem[Fuchs and Cates(2003)]{FuchsFaraday}
M.~Fuchs and M.~E. Cates, \emph{Faraday Discuss.}, 2003, \textbf{123},
  267--286\relax
\mciteBstWouldAddEndPuncttrue
\mciteSetBstMidEndSepPunct{\mcitedefaultmidpunct}
{\mcitedefaultendpunct}{\mcitedefaultseppunct}\relax
\EndOfBibitem
\bibitem[Bra()]{Brader2009note}
In Ref.~\protect\citep{Brader2009}, the stress following step strain was
  evaluated using a linear relation $G(t)=v_\sigma\phi(t)$ instead of the
  quadratic one stated.\relax
\mciteBstWouldAddEndPunctfalse
\mciteSetBstMidEndSepPunct{\mcitedefaultmidpunct}
{}{\mcitedefaultseppunct}\relax
\EndOfBibitem
\bibitem[Venerus \emph{et~al.}(1990)Venerus, Vrentas, and Vrentas]{Venerus1990}
D.~C. Venerus, C.~M. Vrentas and J.~S. Vrentas, \emph{J. Rheol.}, 1990,
  \textbf{34}, 657--683\relax
\mciteBstWouldAddEndPuncttrue
\mciteSetBstMidEndSepPunct{\mcitedefaultmidpunct}
{\mcitedefaultendpunct}{\mcitedefaultseppunct}\relax
\EndOfBibitem
\bibitem[Wagner and Ehrecke(1998)]{Wagner1998}
M.~H. Wagner and P.~Ehrecke, \emph{J. Non-Newt. Fluid Mech.}, 1998,
  \textbf{76}, 183--197\relax
\mciteBstWouldAddEndPuncttrue
\mciteSetBstMidEndSepPunct{\mcitedefaultmidpunct}
{\mcitedefaultendpunct}{\mcitedefaultseppunct}\relax
\EndOfBibitem
\bibitem[Larson and Valesano(1986)]{LarsonValesano}
R.~G. Larson and V.~A. Valesano, \emph{J. Rheol.}, 1986, \textbf{30},
  1093--1108\relax
\mciteBstWouldAddEndPuncttrue
\mciteSetBstMidEndSepPunct{\mcitedefaultmidpunct}
{\mcitedefaultendpunct}{\mcitedefaultseppunct}\relax
\EndOfBibitem
\bibitem[Wagner and Ehrecke(1998)]{WagnerEhrecke}
M.~H. Wagner and P.~Ehrecke, \emph{J. Non-Newt. Fluid Mech.}, 1998,
  \textbf{76}, 183--197\relax
\mciteBstWouldAddEndPuncttrue
\mciteSetBstMidEndSepPunct{\mcitedefaultmidpunct}
{\mcitedefaultendpunct}{\mcitedefaultseppunct}\relax
\EndOfBibitem
\bibitem[Wagner and Stephenson(1979)]{WagnerStephenson}
M.~H. Wagner and S.~E. Stephenson, \emph{J. Rheol.}, 1979, \textbf{23},
  489--504\relax
\mciteBstWouldAddEndPuncttrue
\mciteSetBstMidEndSepPunct{\mcitedefaultmidpunct}
{\mcitedefaultendpunct}{\mcitedefaultseppunct}\relax
\EndOfBibitem
\bibitem[Wang \emph{et~al.}(2006)Wang, Ravindranath, Boukany, Olechnowicz,
  Quirk, Halasa, and Mays]{Wang2006}
S.-Q. Wang, S.~Ravindranath, P.~Boukany, M.~Olechnowicz, R.~P. Quirk, A.~Halasa
  and J.~Mays, \emph{Phys. Rev. Lett.}, 2006, \textbf{97}, 187801\relax
\mciteBstWouldAddEndPuncttrue
\mciteSetBstMidEndSepPunct{\mcitedefaultmidpunct}
{\mcitedefaultendpunct}{\mcitedefaultseppunct}\relax
\EndOfBibitem
\end{mcitethebibliography}
\bibliographystyle{rsc} 
}

\end{document}